\documentclass[prb,aps,twocolumn,superscriptaddress]{revtex4}
\usepackage{graphicx,amsmath,bm}
\usepackage{dcolumn}

\begin{document}

\title{Self-Consistent Approximations for Superconductivity\\
beyond the Bardeen-Cooper-Schrieffer Theory}

\author{Takafumi Kita}
%\email{}
\affiliation{Department of Physics, Hokkaido University,
Sapporo 060-0810, Japan}
\date{\today}

\begin{abstract}
We develop a concise self-consistent perturbation expansion for superconductivity where all the pair processes are naturally incorporated without drawing ``anomalous'' Feynman diagrams.
This simplification results from introducing an interaction vertex that is symmetric in the particle-hole indices besides the ordinary space-spin coordinates. The formalism automatically satisfies conservation laws, includes the Luttinger-Ward theory as the normal-state limit, and reproduces the Bardeen-Cooper-Schrieffer theory as the lowest-order approximation. It enables us to study the thermodynamic, single-particle, two-particle, and dynamical properties of superconductors with competing fluctuations based on a single functional $\Phi[\hat{G}]$ of Green's function $\hat{G}$ in the Nambu space. Specifically, we derive closed equations in the FLEX-S approximation, i.e., the fluctuation exchange approximation for superconductivity with all the pair processes, which contains extra terms besides those in the standard FLEX approximation.
\end{abstract}

\maketitle

\section{Introduction}

The remarkable success of the mean-field Bardeen-Cooper-Schrieffer (BCS) theory in describing classic superconductors \cite{BCS57,Parks69}
may be attributed to the fact that an infinitesimal attraction between electrons suffices to form Cooper pairs responsible for superconductivity.\cite{Cooper56} 
On the other hand, we now have an increasing number of superconductors and Fermi superfluids where relevant interactions are stronger beyond the applicability of the mean-field theory. They include superfluid $^{3}$He at high pressure, \cite{Leggett75,Vollhardt90} heavy-fermion superconductors with competing fluctuations,\cite{Flouquet05,Misra08,Pfleiderer09} high-$T_{c}$ cuprate superconductors,\cite{BK03} and the BCS-BEC (Bose-Einstein condensation) crossover in trapped atomic gases.\cite{GPS08} 
The present paper is devoted to developing a convenient systematic method to describe those systems based on many-body quantum field theory. 

To be specific, we will focus on a two-body interaction and extend the Luttinger-Ward self-consistent perturbation expansion for normal states\cite{LW60} to superconductors in such a way that, given a normal-state approximation, all the pair (i.e., ``anomalous'') processes derivable from it are automatically incorporated. Reproducing the Hartree-Fock theory as the lowest-order approximation, the Luttinger-Ward formalism enables us to include correlation effects systematically and microscopically up to a desired order. Its key ingredient is a functional $\Phi[G]$ of the single-particle Green's function $G$ described by closed skeleton Feynman diagrams, as properly identified and called ``$\Phi$-derivable approximations'' by Baym.\cite{Baym62} Indeed, choosing an approximate $\Phi$ enables us to calculate the whole thermodynamic series ranging from the thermodynamic potential and single-particle Green's function to the two-particle and higher-order Green's functions.\cite{Kita10} To put it another way, $\Phi$ uniquely determines how to resolve the Bogoliubov-Born-Green-Kirkwood-Yvon (BBGKY) hierarchy.\cite{Cercignani88,Kita10} An additional prominent advantage of the formalism is that it automatically satisfies conservation laws, as shown by Baym. \cite{Baym62} Hence, one can also study non-equilibrium phenomena by only changing the imaginary-time Matsubara contour into the real-time Schwinger-Keldysh contour. \cite{Schwinger61,Keldysh64,HJ08,Rammer07,Kita10} 

\begin{figure}[b]
        \begin{center}
                \includegraphics[width=0.9\linewidth]{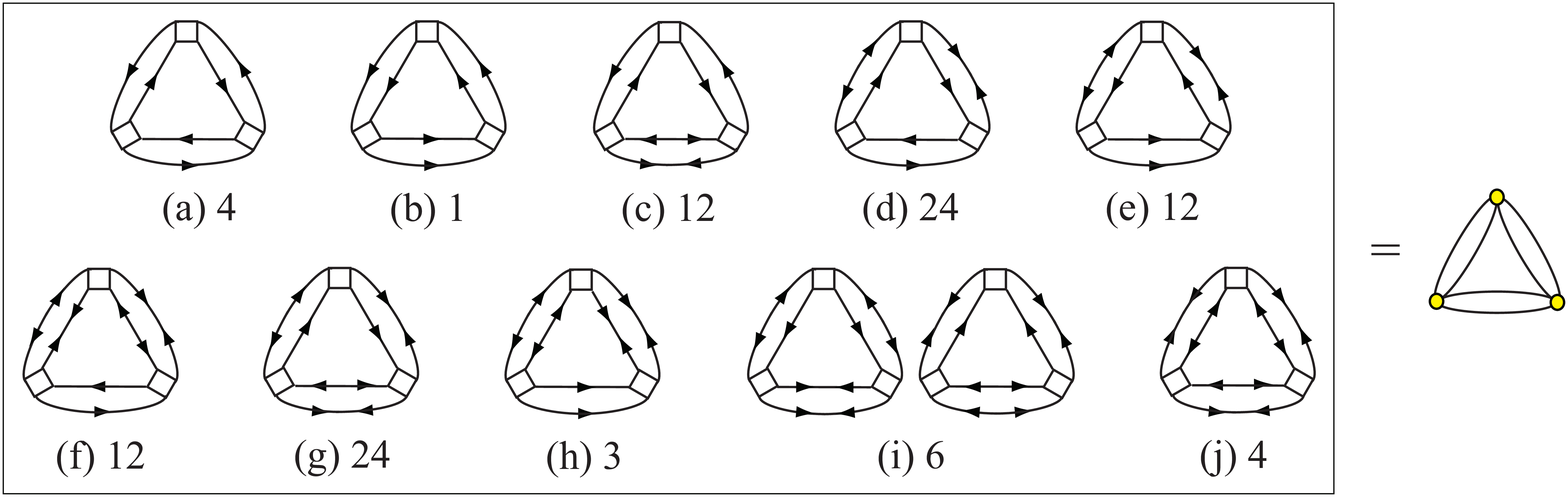}
        \end{center}
        \caption{Third-order diagrams for $\Phi$. A square (a small circle) represents the bare vertex $\Gamma^{(0)}$ ($\underline{\Gamma}^{(0)}$) that is symmetric in the space-spin coordinates (both the space-spin coordinates and particle-hole indices). The number below each of diagrams (a)-(j) denotes its relative weight. \label{Fig1}}
\end{figure}

It should be noted that a formal extension of the Luttinger-Ward theory to superconductors was performed by de Dominicis and Martin  already in 1964.\cite{dDM64} However, they did not present any practical method of how to efficiently collect pair processes with anomalous Green's functions, whose number grows exponentially in comparison with normal diagrams as we proceed to higher orders. For example, (a)-(j) in Fig.\ \ref{Fig1} exhaust the third-order diagrams for $\Phi$ in terms of the symmetrized vertex $\Gamma^{(0)}$ of Abrikosov {\em et al}.,\cite{AGD63} where anomalous diagrams (c)-(j) are already four times as large in number as normal diagrams (a) and (b).

Here, we develop a perturbation expansion of gathering all the pair processes systematically and concisely without drawing anomalous diagrams for both the singlet and triplet pairings or their mixtures. This is made possible by introducing a bare interaction vertex $\underline{\Gamma}^{(0)}$ that is further symmetrized in the particle-hole (i.e., creation-annihilation or ``charge'') indices. Thus, processes (a)-(j) in Fig.\ \ref{Fig1} can be represented by a single diagram on the right-hand side with a definite analytic expression, i.e., eq.\ (\ref{Phi_3}) below, which reproduces their weights correctly. The key functional $\Phi$ is defined as a closed skeleton expansion in terms of the matrix Green's function $\hat{G}$ in the Nambu space, which may be approximated by some selected terms from the whole series in practical calculations. Besides the Dyson-Gor'kov equation for $\hat{G}$ and an expression of the thermodynamic potential, we also derive the Bethe-Salpeter equation for the two-particle Green's function that can be solved based solely on $\Phi$. The formalism enables us to study not only superconductivity itself but also competing effects of charge, spin, and superconducting fluctuations or orders simultaneously through the matrix structure of $\underline{\Gamma}^{(0)}$. This last point may be regarded as a definite advantage of the present approach over the one based on the Stratonovich-Hubbard transformation,\cite{Hubbard59} where only a single fluctuation relevant to the auxiliary field may be taken into account after some standard approximation.

Using the formalism, we also extend the fluctuation exchange (FLEX) approximation originally developed for normal states\cite{BS89,BSW89,BW91} so as to incorporate all the pair processes for the spin-singlet pairing. The FLEX approximation has been successful in describing anomalous normal-state transport phenomena of high-$T_{c}$ superconductors as well as some organic and heavy-fermion superconductors.\cite{Kontani08} Its extension to superconductivity as described above, which has not been performed yet and will be called ``FLEX-S approximation'' below, will be useful for understanding those superconductors quantitatively.
This will be carried out in a transparent and concise manner in terms of $\underline{\Gamma}^{(0)}$ in contrast to the original derivation for normal states.\cite{BS89,BSW89,BW91}

Naturally, considerable efforts have been made to go beyond the BCS theory.
Eliashberg incorporated the electron-phonon interaction explicitly as the source of the attraction to derive ``strong-coupling'' equations of superconductivity within the second-order perturbation in terms of the renormalized electron Green's function and bare phonon propagator.\cite{Eliashberg60} Leggett seriously considered interactions between particles other than the pairing part to develop a theory of superfluid Fermi liquids at low temperatures.\cite{Leggett65} 
Anderson and Brinkman \cite{AB73} presented the idea of ``feedback effects,''  i.e., a change of the pairing interaction through the superfluid transition,
to understand the A phase of superfluid $^3$He that is realized against the mean-field Balian-Werthamer theory.\cite{BW63}
The idea was elaborated by Brinkman {\em et al}.\ \cite{BSA74} and Kuroda,\cite{Kuroda75} who both incorporated the first few anomalous processes into the normal-state pairing interaction composed of the particle-hole ``paramagnon'' diagrams\cite{LF71} and evaluated the extra free energy as a power series in the $p$-wave energy gaps. Subsequently, Tewordt \cite{Tewordt74} refined this approach into a self-consistent $\Phi$-derivable approximation, where he included those pair processes that can be obtained from the normal particle-hole diagrams by successively replacing a pair of normal Green's functions connecting adjacent vertices by a pair of anomalous Green's functions; relevant diagrams in the third order are (a), (c), (f), and (j) of Fig.\ \ref{Fig1}.
Tewordt and coworkers\cite{Tewordt93,Tewordt95} later applied the formalism to the two-dimensional Hubbard model to clarify quasiparticle and spin excitations of high-$T_c$ superconductors.
The same formalism was used by other groups for the two-dimensional Hubbard model\cite{PB94,MS94,GLSB96} and $d$-$p$ model.\cite{LB93,KFY97,TM98} 
On the other hand, Eagles\cite{Eagles69} and Leggett\cite{Leggett80} discussed the BCS-BEC crossover of the isotropic spin-singlet pairing at zero temperature by combining the BCS gap equation with the equation for the chemical potential. The transition temperature of the BCS-BEC crossover problem was calculated subsequently by Nozi\`eres and Schmitt-Rink\cite{NSR85} based on the Thouless criterion.\cite{Thouless60} 
Haussmann and coworkers\cite{Haussmann93,Haussmann07} studied this problem with an improved $\Phi$-derivable approximation to include interaction effects among unpaired and paired fermions. Specifically, their $\Phi$ consists of the normal particle-particle ladder diagrams appropriate for low-density systems, plus those anomalous processes obtained from the formers by successively replacing two pairs of normal Green's functions connecting adjacent vertices by two pairs of anomalous Green's functions;
relevant diagrams in the third order are (b) and (h) of Fig.\ 1, disregarding (e) and (g) which may be important even within the ladder approximation.
We should also mention an alternative approach on the issue where some effective pairing interaction is used to solve the Eliashberg equations.\cite{CPS03} However, there is an ambiguity as to how to construct the effective interaction.
Similarly, the FLEX approximation has sometimes been augmented rather phenomenologically to explain the pseudogap phenomena observed in underdoped cuprate superconductors.\cite{DMT97,Yanase01}
To be specific, the ``FLEX+T-matrix'' approximation\cite{Yanase01} consists of (i) identifying the pairing interaction obtained by the FLEX approximation as the ``bare'' interaction in the Thouless criterion 
and (ii) incorporating the contribution of the two-particle ``superconducting fluctuations'' additionally into the single-particle self-energy. 
However, the procedure may contain some double counting of elementary processes and has yet to be examined on a firm microscopic basis.
Note in this context that the Thouless criterion was originally derived by the particle-particle ladder approximation in the bare perturbation expansion for a two-particle Green's function,\cite{Thouless60} reproducing the same $T_c$ equation as the mean-field BCS theory.
Thus, one may be convinced that we still do not have practical methods of incorporating all the anomalous processes that are naturally present below $T_c$. 

This paper is organized as follows. In \S 2, we develop a basic formalism to calculate the thermodynamic potential, single-particle Green's function, and two-particle Green's function of superconductivity self-consistently based on a single functional $\Phi[\hat{G}]$, whose expression may be obtained by a concise perturbation expansion in terms of $\underline{\Gamma}^{(0)}$. We carry it out in the coordinate representation so that general inhomogeneous systems 
may be handled.
In \S 3, we focus on homogeneous systems and transform the results of \S 2 into the momentum-``energy'' representation.
Closed equations in the FLEX-S approximation are also derived. Section 4 presents a brief summary to indicate main results.
The whole contents of the present paper may be regarded as the Fermi-superfluid counterpart of the self-consistent perturbation expansion recently developed for  BEC.\cite{Kita09,Kita10b, Kita11}
We set $\hbar=k_{\rm B}=1$ throughout with $k_{\rm B}$ the Boltzmann constant.

\section{Basic formalism}

The system we consider consists of identical particles with mass $m$ and spin $1/2$ described by the Hamiltonian
\begin{equation}
H=H_{0}+H_{\rm int},
\label{Hamil}
\end{equation}
with
\begin{subequations}
\label{Hamil2}
\begin{equation}
H_{0}=\int {\rm d}\xi_{1} \psi^{\dagger}(\xi_{1})K_{1}\psi(\xi_{1}),
\label{H_0}
\end{equation}
\begin{equation}
H_{\rm int}=\frac{1}{2}\int {\rm d}\xi_{1}\int {\rm d}\xi_{2}\,\psi^{\dagger}(\xi_{1})
\psi^{\dagger}(\xi_{2}) V({\bm r}_{1}-{\bm r}_{2})
 \psi(\xi_{2})\psi(\xi_{1}) .
\label{H_int}
\end{equation}
\end{subequations}
Here 
$\xi_{1}\equiv ({\bm r}_{1},\alpha_{1})$ with ${\bm r}_{1}$ and $\alpha_{1}$ denoting the space and spin coordinates, respectively,\cite{AGD63}
$\psi^{\dagger}$ and $\psi$ are the creation and annihilation operators of the fermion field,  respectively,\cite{AGD63}
$K_{1}\equiv -{\hbar^{2}\nabla_{1}^{2}}/{2m}-\mu$
with $\mu$ the chemical potential,
and $V$ is the interaction potential. 
Though disregarded here, the effect of a lattice or trap potential may be included easily in $K_{1}$.
The formulation in this section will be carried out in the coordinate representation so that general inhomogeneous systems can be handled.
It is also applicable to lattice models such as the Hubbard model, for which every integration over ${\bm r}$ above should be replaced by a summation
over the lattice sites.

Let us introduce the Heisenberg representations of the field operators in the Matsubara formalism by \cite{AGD63}
\begin{equation}
\left\{
\begin{array}{l}
\vspace{1mm}
\psi^{\rm H}_{1}(1) \equiv e^{\tau_{1}H}\psi(\xi_{1})e^{-\tau_{1}H} \\
\psi^{\rm H}_{2}(1) \equiv e^{\tau_{1}H}\psi^{\dagger}(\xi_{1})e^{-\tau_{1}H}
\end{array} \right. ,
\end{equation}
where argument $1$ in the round brackets is defined by $1\equiv (\xi_{1},\tau_{1})$ and variable
$\tau_{1}$ lies in $0 \leq \tau_{1} \leq \beta\equiv 1/T$ with 
$T$ denoting the temperature.
The operators $\psi^{\rm H}_{1}(1)$ and
$\psi^{\rm H}_{2}(1)$ are denoted as $\tilde{\psi}(1)$ and $\tilde{\bar{\psi}}(1)$ by Abrikosov {\em et al}.\ \cite{AGD63}, respectively;
distinguishing them by the subscript $i=1,2$ (i.e., the particle-hole or ``charge'' index) as above enables us to simplify the notation and formulation considerably, as seen below.

\subsection{Green's function, Dyson-Gor'kov equation, and thermodynamic potential\label{subsec:TP}}

Using $\psi^{\rm H}_{i}(1)$, we define Green's functions by
\begin{equation}
G_{ij}(1,2)\equiv - \bigl< T_{\tau} \psi^{\rm H}_{i}(1)\psi^{\rm H}_{3-j}(2)\bigr> ,
\label{G_ij}
\end{equation}
where $T_{\tau}$ is the ``time''-ordering operator and $\langle\cdots\rangle$ denotes the grand-canonical average in terms of $H$.\cite{AGD63} The standard normal-state Green's function $G$ corresponds to $G_{11}$. The elements satisfy
\begin{equation}
G_{ij}(1,2)= -G_{3-j,3-i}(2,1)= G_{ji}^{*}(\xi_{2}\tau_{1},\xi_{1}\tau_{2}),
\label{G_ij-symm}
\end{equation}
with the superscript $^{*}$ denoting the complex conjugate.
The Nambu matrix $\hat{G}\equiv (G_{ij})$ is written explicitly as
\begin{equation}
\hat{G}(1,2)\equiv 
\begin{bmatrix}
G_{11}(1,2) & G_{12}(1,2)
\\
G_{21}(1,2) & G_{22}(1,2)
\end{bmatrix} .
\label{hatG}
\end{equation}
It obeys the Dyson-Gor'kov equation
\begin{equation}
\bigl[\hat{G}_{0}^{-1}(1,\bar{3})-\hat{\Sigma}(1,\bar{3})\bigr] \hat{G}(\bar{3},2)=\hat{\sigma}_{0} \delta(1,2),
\label{DG}
\end{equation}
where $\hat{\sigma}_{0}$ is the $2\times 2$ unit matrix,
$\delta(1,2)\equiv \delta(\tau_{1}-\tau_{2})\delta(\xi_{1}-\xi_{2})$, and integrations over barred arguments are implied.
The quantity $\hat{G}_{0}^{-1}$ is defined by
\begin{equation}
\hat{G}_{0}^{-1}(1,2)\equiv \left(-\hat{\sigma}_{0}\frac{\partial}{\partial \tau_{1}}
-\hat{\sigma}_{3}K_{1}\right)\delta(1,2) ,
\label{G_0^-1(1,2)}
\end{equation}
where $\hat{\sigma}_{3}$ is the third Pauli matrix.
Finally, $\hat{\Sigma}$ in eq.\ (\ref{DG}) denotes the self-energy matrix
\begin{equation}
\hat{\Sigma}(1,2)\equiv 
\begin{bmatrix}
\Sigma_{11}(1,2) & \Sigma_{12}(1,2)
\\
\Sigma_{21}(1,2) & \Sigma_{22}(1,2)
\end{bmatrix} .
\label{hatSigma}
\end{equation}
It follows from eqs.\ (\ref{G_ij-symm}), (\ref{DG}), and (\ref{G_0^-1(1,2)}) that $\hat{\Sigma}$ satisfies the same symmetry relations as $\hat{G}$ in eq.\ (\ref{G_ij-symm}).

As shown by de Dominicis and Martin,\cite{dDM64} we can express 
the thermodynamic potential  $\Omega\equiv -\beta^{-1}\ln {\rm Tr}\, {\rm e}^{-\beta H}$
as a functional of $\hat{G}$, i.e., $\Omega=\Omega[\hat{G}]$, so as to satisfy
\begin{equation}
\frac{\delta \Omega}{\delta G_{ji}(2,1)}=0.
\label{dOmega/dG=0}
\end{equation}
Let us write $\Omega[\hat{G}]$ as
\begin{equation}
\Omega = -\frac{1}{2\beta}{\rm Tr}\bigl[ \ln \bigl(-\hat{G}_{0}^{-1}+\hat{\Sigma}\bigr)+\hat{\Sigma}\hat{G}\bigr]+\Phi ,
\label{Omega}
\end{equation}
where $\hat{\Sigma}=\hat{\Sigma}[\hat{G}]$, $\Phi=\Phi[\hat{G}]$, and the operator ${\rm Tr}$ for the Nambu matrices is defined by
\begin{equation}
{\rm Tr}\,\hat{A}\equiv 
 A_{11}(\bar{1},\bar{1}_{+}) + A_{22}(\bar{1}_{+},\bar{1})
\label{Tr-def}
\end{equation}
with the subscript of $1_+$ denoting an extra infinitesimal positive constant in $\tau_{1}$ to 
place creation operators to the left of annihilation ones for the equal-time average.
Using eqs.\ (\ref{DG}) and (\ref{Omega}), one may show easily that eq.\ (\ref{dOmega/dG=0}) is transformed into a relation between 
$\hat{\Sigma}$ and $\Phi$ as
\begin{equation}
\Sigma_{ij}(1,2)=2\beta\frac{\delta \Phi}{\delta G_{ji}(2,1)} .
\label{Sigma-Phi}
\end{equation}
Equation (\ref{Omega}) with eqs.\ (\ref{DG}) and (\ref{Sigma-Phi}) forms an extension of the normal-state Luttinger-Ward functional\cite{LW60} to superconductors.
Note that we have given $\Omega$ as a functional of $\hat{G}$ following Baym\cite{Baym62} instead of $\hat{\Sigma}$ in the original treatment.\cite{LW60}

It follows from eqs.\ (\ref{DG}), (\ref{Omega}), and (\ref{Sigma-Phi}) that we can calculate $\hat{G}$ and $\Omega$ self-consistently once $\Phi=\Phi[\hat{G}]$ is given explicitly. The particle number $N$ is obtained by
\begin{equation}
N=\frac{1}{2}{\rm Tr}\,\hat{G} \hat{\sigma}_{3},
\end{equation}
which may be used to change an independent variable from $\mu$ to $N$ by the Legendre transformation $F\equiv \Omega+\mu N$.

\subsection{S-matrix and functional $\Phi$}

\begin{figure}[t]
        \begin{center}
                \includegraphics[height=10mm]{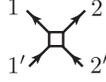}
        \end{center}
        \caption{Bare interaction vertex $\Gamma^{(0)}$ that is symmetric in a pair of outgoing (incoming) lines. \label{Fig2}}
\end{figure}

Perturbation expansions of $\Omega$ and $\hat{G}$ in terms of $H_{\rm int}$ in eq.\ (\ref{H_int})
can be carried out conveniently with the S-matrix\cite{AGD63}
\begin{equation}
{\cal S} \equiv T_{\tau} \exp\biggl[ -\frac{1}{4} 
\Gamma^{(0)}(\bar{1}\bar{1}',\bar{2}\bar{2}')
\psi_{2}(\bar{1})\psi_{2}(\bar{2})\psi_{1}(\bar{2}')\psi_{1}(\bar{1}')\biggr] .
\label{S-matrix}
\end{equation}
Here $\psi_{i}(1)$ ($i \!=\! 1,2$) are the interaction representations of the field operators given explicitly by $\psi_{1}(1) \equiv e^{\tau_{1}H_{0}}\psi(\xi_{1})e^{-\tau_{1}H_{0}}$ and
$\psi_{2}(1) \equiv e^{\tau_{1}H_{0}}\psi^{\dagger}(\xi_{1}) e^{-\tau_{1}H_{0}}$,
and $\Gamma^{(0)}$ denotes the symmetrized bare vertex\cite{AGD63}
\begin{eqnarray}
&&\hspace{-14mm}
\Gamma^{(0)}(11',22')\equiv V({\bm r}_{1}-{\bm r}_{2})\delta(\tau_{1}-\tau_{2})
\nonumber \\
&&\hspace{10mm}\times
 [\delta(1,1')\delta(2,2')-\delta(1,2')\delta(2,1')] ,
\label{Gamma^(0)}
\end{eqnarray}
satisfying
$\Gamma^{(0)}(11',22')\!=\!\Gamma^{(0)}(22',11')\!=\! \Gamma^{(0)}(1'1,2'2)\!=\! -\Gamma^{(0)}(12',21')$.
The interaction in eq.\ (\ref{S-matrix}) can be expressed graphically as Fig.\ \ref{Fig2}.

Now, the key functional $\Phi$ in eq.\ (\ref{Omega}) is given exactly in terms of eq.\ (\ref{S-matrix}) as
\begin{equation}
\Phi[\hat{G}] = -\frac{<{\cal S}>_{0{\rm c}}-1}{\beta}\biggr|_{{\rm skeleton},\hat{G}_{0}\rightarrow \hat{G}},
\label{Phi}
\end{equation}
where subscript $_{0}$ denotes the thermodynamic average with $H_{0}$ and another subscript $_{\rm c}$ 
implies retaining only connected Feynman diagrams.
The right-hand side may be expressed graphically by the skeleton diagrams (i.e., diagrams without self-energy insertions) in the perturbation expansion for $\Omega$ with $\hat{G}_{0}$ replaced by $\hat{G}$.\cite{LW60} Approximating eq.\ (\ref{Phi}) by a few terms or some partial series, we can construct a self-consistent 
approximation, as already noted in the last paragraph of \S \ref{subsec:TP}.
Equation (\ref{Phi}) was proved by Luttinger and Ward for normal states.\cite{LW60,Kita10} However, it also holds true for superconductors by including all the pair processes.

\begin{figure}[t]
        \begin{center}
                \includegraphics[height=17mm]{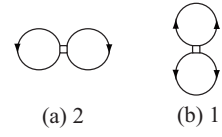}
        \end{center}
        \caption{First-order diagrams for $\Phi$. The number below each diagram denotes its relative weight.\label{Fig3}}
\end{figure}

The Feynman rules for the perturbation expansion of eq.\ (\ref{Phi}) with eq.\ (\ref{S-matrix}) was given by Abrikosov {\em et al}.\ \cite{AGD63} for normal states.
Following them, however, we need to calculate numerical factors from a separate expansion in terms of $V$.
This may be the reason why the expansion with $\Gamma^{(0)}$ has not been used widely.
It is shown in Appendix\ref{appendix:Feynman} that the rules can be simplified considerably so that one may perform it directly in terms of $\Gamma^{(0)}$,
even for superconductors.
For example, the first-order contribution to $\Phi$ is graphically given by Fig.\ \ref{Fig3}, 
which can be expressed analytically as
\begin{eqnarray}
&&\hspace{-10mm}
\Phi_{1}=\frac{1}{4\beta}\Gamma^{(0)}(\bar{1}\bar{1}',\bar{2}\bar{2}')\bigl[2G(\bar{1}',\bar{1})G(\bar{2}',\bar{2})
\nonumber \\
&&\hspace{-1mm}
+\bar{F}(\bar{1},\bar{2})F(\bar{1}',\bar{2}')\bigr],
\label{Phi_1}
\end{eqnarray}
where $G$, $F$, and $\bar{F}$ are defined together with $\bar{G}$ by
\begin{subequations}
\label{GF-def}
\begin{eqnarray}
&&\hspace{-10mm}
G(1,2)\equiv \frac{G_{11}(1,2)-G_{22}(2,1)}{2}\equiv \bar{G}(2,1),
\\
&&\hspace{-10mm}
F(1,2)\equiv \frac{G_{12}(1,2)-G_{12}(2,1)}{2},
\\
&&\hspace{-10mm}
\bar{F}(1,2)\equiv -\frac{G_{21}(1,2)-G_{21}(2,1)}{2}.
\end{eqnarray}
\end{subequations}
With eq.\ (\ref{GF-def}), we have incorporated the first symmetry of eq.\ (\ref{G_ij-symm}) manifestly into $\Phi_{1}$
for the purpose of deriving $\underline{\Gamma}^{(0)}$ below.

\subsection{Concise perturbation expansion for $\Phi$}

The expansion with $\Gamma^{(0)}$ is still cumbersome for collecting anomalous diagrams
in superconductivity, however, as seen from Fig.\ \ref{Fig1}.
Hence, we further modify the expression of eq.\ (\ref{S-matrix}) so that the asymmetry between the incoming and outgoing lines
in Fig.\ \ref{Fig2} is removed. 
To be specific, we introduce a bare vertex $\Gamma^{(0)}_{ii',jj'}(11',22')$ from eq.\ (\ref{Phi_1}) by
\begin{subequations}
\label{hatGamma^(0)}
\begin{equation}
\Gamma^{(0)}_{ii',jj'}(11',22')\equiv 2\beta\frac{\delta^{2}\Phi_{1}}{\delta G_{i'i}(1',1)G_{j'j}(2',2)}.
\label{hatGamma^(0)-1}
\end{equation}
A straightforward calculation using the symmetry of eq.\ (\ref{Gamma^(0)}) shows that results of all the above differentiations may be summarized as
\begin{eqnarray}
&&\hspace{-12mm}
\Gamma^{(0)}_{ii',jj'}(11',22')=\frac{\delta_{ij}}{2}\delta_{ii'}\delta_{jj'}\Gamma^{(0)}(11',22')
\nonumber \\
&&\hspace{16mm}
-\frac{\delta_{i,3-j}}{2}\delta_{ii'}\delta_{jj'}\Gamma^{(0)}(11',2'2)
\nonumber \\
&&\hspace{16mm}
-\frac{\delta_{i,3-j}}{2}\delta_{ij'}\delta_{ji'}\Gamma^{(0)}(12,1'2') ,
\label{hatGamma^(0)-2}
\end{eqnarray}
\end{subequations}
which satisfies
\begin{eqnarray}
&&\hspace{-4mm}
\Gamma^{(0)}_{ii',jj'}(11',22')=\Gamma^{(0)}_{jj',ii'}(22',11')=-\Gamma^{(0)}_{ij',ji'}(12',21')
\nonumber \\
&&\hspace{-8mm}
=-\Gamma^{(0)}_{3-i',3-i,jj'}(1'1,22')=-\Gamma^{(0)}_{3-j',i',j,3-i}(2'1',21).
\label{hatGamma-symm}
\end{eqnarray}
Now, we can express eq.\ (\ref{S-matrix}) alternatively as
\begin{eqnarray}
&&\hspace{-10mm}
{\cal S}= T_{\tau} \exp\biggl[-\frac{1}{12}\Gamma^{(0)}_{\bar{i}\bar{i}',\bar{j}\bar{j}'}(\bar{1}\bar{1}',\bar{2}\bar{2}')
\nonumber \\
&&\hspace{-3mm}\times
{\cal N}\psi_{\bar{i}'}(\bar{1}')\psi_{3-\bar{i}}(\bar{1})\psi_{\bar{j}'}(\bar{2}')\psi_{3-\bar{j}}(\bar{2})\biggr] ,
\label{S-matrix2}
\end{eqnarray}
where ${\cal N}$ denotes the normal-ordering operator of placing creation operators to the left of annihilation operators
with a sign change per every permutation of a pair of adjacent field operators.\cite{AGD63}
The equivalence between eqs.\ (\ref{S-matrix}) and (\ref{S-matrix2}) may be checked easily by substituting eq.\ (\ref{hatGamma^(0)-2}) into the latter
and writing the resultant expression without the ${\cal N}$ operator.
Note that ${\cal N}$ is only relevant to equal-time averages of the first order in the perturbation expansion.

\begin{figure}[t]
        \begin{center}
                \includegraphics[height=10mm]{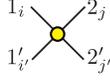}
        \end{center}
        \caption{Bare interaction vertex $\Gamma^{(0)}_{ii',jj'}(11',22')$ that is symmetric in four external lines. \label{Fig4}}
\end{figure}

\begin{figure}[t]
        \begin{center}
                \includegraphics[height=14mm]{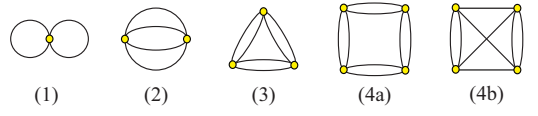}
        \end{center}
        \caption{First- to forth-order diagrams for $\Phi$. \label{Fig5}}
\end{figure}

The interaction in eq.\ (\ref{S-matrix2}) may be expressed graphically as Fig.\ \ref{Fig4}, 
which is symmetric in the four external lines.
Using eq.\ (\ref{S-matrix2}), we can perform the perturbation expansion for $\Phi$ (and also for any other quantities) concisely in such a way that all the anomalous processes are incorporated automatically.  
Figure \ref{Fig5} enumerates first- to forth-order diagrams for $\Phi$.
Thus, each of the first- to third-order contributions is exhausted by a single diagram, showing manifestly the advantage of using eq.\ (\ref{S-matrix2}).
Moreover, the symmetry of eq.\ (\ref{hatGamma-symm}) brings the simplification that every possible Wick decomposition for a distinct diagram 
yields the same contribution.
Hence, we only need to consider a convenient decomposition (i.e., connection of vertices)  for each distinct diagram and multiply the result by the number of possible connections,
which may be calculated easily based on a combinatorial consideration.

Now, the Feynman rules for the expansion of eq.\ (\ref{Phi}) with eq.\ (\ref{S-matrix2}) are summarized as follows.
\begin{enumerate}
\item[(a)] Draw all the $n$th-order closed skeleton diagrams that are topologically distinct. 
For each such diagram, associate the factor $(-1)^{n+1}/ n!12^{n}\beta$.

\item[(b)] For each small circle, associate $\Gamma^{(0)}_{\bar{i}\bar{i}',\bar{j}\bar{j}'}(\bar{\lambda}\bar{\lambda}',\bar{\nu}\bar{\nu}')$.

\item[(c)] Identify the number $C_{n\alpha}$ of possible connections of vertices for diagram $\alpha$ under consideration.

\item[(d)] Consider a specific connection of vertices for diagram $\alpha$ where every primed argument is linked with an unprimed argument, 
and associate $G_{\bar{i}'\bar{j}}(\bar{\lambda}',\bar{\eta})$ for each line connecting $\bar{\lambda}_{\bar{i}'}'$ and $\bar{\eta}_{\bar{j}}$.

\item[(e)] Identify the number $\ell_{n\alpha}$ of permutations to realize the connection. 
With the choice in (d), we may find it by replacing every vertex as
$\Gamma^{(0)}_{ii',jj'}(11',22')\rightarrow \delta_{i1}\delta_{i'1}\delta_{j1}\delta_{j'1}\delta(1,1')\delta(2,2') V({\bm r}_{1}-{\bm r}_{2})\delta(\tau_{1}-\tau_{2})$
and equating $\ell_{n\alpha}$ with the number of closed particle loops in the resultant normal diagram with $V$.\cite{LW60}

\item[(f)] Multiply the expression by $(-1)^{\ell_{n\alpha}}C_{n\alpha}$.

\item[(g)] When calculating the two-particle irreducible vertex $\underline{\Gamma}^{({\rm ir})}$ by eq.\ (\ref{Gamma^(ir)}) below, replace every Green's function as
\begin{equation}
G_{ij}(1,2)\rightarrow \tilde{G}_{ij}(1,2)\equiv \frac{G_{ij}(1,2)-G_{3-j,3-i}(2,1)}{2},
\label{tG_ij}
\end{equation}
so as to incorporate the first symmetry of eq.\ (\ref{G_ij-symm}) manifestly in $\underline{\Gamma}^{({\rm ir})}$.

\end{enumerate}

\noindent 
Rule (g) is relevant only to two-particle and higher-order Green's functions.

As an example of rule (c) above, let us consider diagram (3) in Fig.\ \ref{Fig5}. Its Wick decomposition may proceed as follows:
Start from an arbitrary operator $\psi_{i}$ (labelled $a_{1}$ operator in $H_{\rm int}^{a}$) and find its partner (labelled $b_{1}$ operator in $H_{\rm int}^{b}$) from 
$2\times 4$ possible candidates; to connect another line between $H_{\rm int}^{a}$ and $H_{\rm int}^{b}$, pick out a pair of operators $a_{2}$ and $b_{2}$ from 
$3\times 3$ possible choices; to connect a pair of lines between $H_{\rm int}^{b}$ and $H_{\rm int}^{c}$, find the partners of $b_{3}$ and $b_{4}$ (labelled $c_{1}$ and $c_{2}$ operators, respectively) from $4\times 3$ possible choices in $H_{\rm int}^{c}$; finally, select one of the $2$ possibilities to  
connect a pair of lines between ($c_{3}$,$c_{4}$) and ($a_{3}$,$a_{4}$). Thus, the number $C_{3}$ is obtained as
$$
C_{3}=(2\cdot 4\cdot 3^{2})\cdot (4\cdot 3)\cdot 2 =2!4^{2}3^{3}2.
$$
Now, rules (a)-(f) enable us to write down an analytic expression for $\Phi_{3}$ as
\begin{eqnarray}
&&\hspace{-10mm} 
\Phi_{3} =
\frac{(-1)^{4+\ell_{3}}C_{3}}{3!12^{3}\beta}
\Gamma^{(0)}_{\bar{i}\bar{i}',\bar{j}\bar{j}'}(\bar{1}\bar{1}',\bar{2}\bar{2}')
G_{\bar{j}'\bar{k}}(\bar{2}',\bar{3})G_{\bar{k}'\bar{j}}(\bar{3}',\bar{2})
\nonumber \\
&& \hspace{-1mm} \times 
\Gamma^{(0)}_{\bar{k}\bar{k}',\bar{l}\bar{l}'}(\bar{3}\bar{3}',\bar{4}\bar{4}')
G_{\bar{l}'\bar{m}}(\bar{4}',\bar{5})G_{\bar{m}'\bar{l}}(\bar{5}',\bar{4})
\nonumber \\
&& \hspace{-1mm}  \times 
\Gamma^{(0)}_{\bar{m}\bar{m}',\bar{n}\bar{n}'}(\bar{5}\bar{5}',\bar{6}\bar{6}')
G_{\bar{n}'\bar{i}}(\bar{6}',\bar{1})G_{\bar{i}'\bar{n}}(\bar{1}',\bar{6}),
\label{Phi_3-0}
\end{eqnarray}
with $\ell_{3}=3$ from rule (e) for this connection.
An elementary calculation using eq.\ (\ref{hatGamma^(0)-2}) shows that eq.\ (\ref{Phi_3-0}) correctly reproduces the weights of 
diagrams (a)-(j) in Fig.\ \ref{Fig1}.
It is convenient at this stage to introduce matrices $\underline{\Gamma}^{(0)}$ and $\underline{\chi}^{(0)}$ by
\begin{subequations}
\label{uGamma-uchi}
\begin{eqnarray}
&&\hspace{-10mm}
\langle 11'_{ii'}|\underline{\Gamma}^{(0)}|22'_{jj'}\rangle \equiv
{\Gamma}^{(0)}_{ii',j'j}(11',2'2),
\label{uGamma^(0)}
\\
&&\hspace{-10mm}
\langle 11'_{ii'}|\underline{\chi}^{(0)}|22'_{jj'}\rangle \equiv
-G_{ij}(1,2)G_{j'i'}(2',1').
\label{chi^(0)}
\end{eqnarray}
\end{subequations}
Using them, we can express eq.\ (\ref{Phi_3-0}) concisely as
\begin{subequations}
\label{Phi_1-4}
\begin{equation}
\Phi_{3} = \frac{1}{6\beta}{\rm Tr}\,\bigl(\underline{\Gamma}^{(0)}\underline{\chi}^{(0)}\bigr)^{\! 3},
\label{Phi_3}
\end{equation}
where the basis of Tr is given by the bracket vectors in eq.\ (\ref{uGamma-uchi})

The above consideration may be extended to the other diagrams in Fig.\ \ref{Fig5},
where the numbers of rule (c) are identified as  $C_{1}=3$, $C_{2}=4!$, $C_{4a}=3!4^{3}3^{4}2$, and 
$C_{4b}=(_4 C_2)^{4}3\cdot 2^{2}2^{3}=3^{5}2^{9}=3!4^{4}3^{4}$.
We thereby obtain 
\begin{equation}
\Phi_{1}=\frac{1}{4\beta}{\rm Tr}\,\underline{\Gamma}^{(0)}\underline{\chi}^{(0)},
\label{Phi_1-2}
\end{equation}
\begin{equation}
\Phi_{2}=-\frac{1}{12\beta}{\rm Tr}\,\bigl(\underline{\Gamma}^{(0)}\underline{\chi}^{(0)}\bigr)^{\! 2},
\end{equation}
\begin{equation}
\Phi_{4a}=-\frac{1}{8\beta}{\rm Tr}\,\bigl(\underline{\Gamma}^{(0)}\underline{\chi}^{(0)}\bigr)^{\! 4},
\label{Phi_4a}
\end{equation}
\begin{eqnarray}
&&\hspace{-15mm} 
\Phi_{4b} =
\frac{1}{4\beta}
\Gamma^{(0)}_{\bar{i}\bar{i}',\bar{j}\bar{j}'}(\bar{1}\bar{1}',\bar{2}\bar{2}')
\Gamma^{(0)}_{\bar{k}\bar{k}',\bar{l}\bar{l}'}(\bar{3}\bar{3}',\bar{4}\bar{4}')
\nonumber \\
&& \hspace{-4.5mm} \times 
\Gamma^{(0)}_{\bar{m}\bar{m}',\bar{n}\bar{n}'}(\bar{5}\bar{5}',\bar{6}\bar{6}')
\Gamma^{(0)}_{\bar{p}\bar{p}',\bar{q}\bar{q}'}(\bar{7}\bar{7}',\bar{8}\bar{8}')
\nonumber \\
&& \hspace{-4.5mm} \times 
G_{\bar{j}'\bar{k}}(\bar{2}',\bar{3})G_{\bar{k}'\bar{j}}(\bar{3}',\bar{2})
G_{\bar{n}'\bar{p}}(\bar{6}',\bar{7})G_{\bar{p}'\bar{n}}(\bar{7}',\bar{6})
\nonumber \\
&& \hspace{-4.5mm} \times 
G_{\bar{i}'\bar{q}}(\bar{1}',\bar{8})G_{\bar{q}'\bar{l}}(\bar{8}',\bar{4})G_{\bar{l}'\bar{m}}(\bar{4}',\bar{5})G_{\bar{m}'\bar{i}}(\bar{5}',\bar{1}).
\label{Phi_4b}
\end{eqnarray}
\end{subequations}
The equivalence between eqs.\ (\ref{Phi_1}) and (\ref{Phi_1-2}) may be checked easily by using eqs.\ (\ref{GF-def}), (\ref{hatGamma^(0)-2}),
and (\ref{uGamma-uchi}).
Now, one may be convinced that eq.\ (\ref{S-matrix2}) considerably simplifies the perturbation expansion of superconductivity.

\subsection{Self-consistent approximations}

As already mentioned, eq.\ (\ref{Phi}) forms a convenient basis for systematic approximations for superconductivity.
To be specific, Green's function and the self-energy are determined self-consistently by eqs.\ (\ref{DG}) and (\ref{Sigma-Phi}) 
based on some approximate $\Phi$.
The lowest-order approximation corresponds to the choice $\Phi\approx \Phi_{1}$ from the series of eq.\ (\ref{Phi_1-4}), which is exactly the mean-field BCS theory.\cite{BCS57,Parks69}
It may be improved by including the next-order term as $\Phi\approx \Phi_{1}+\Phi_{2}$, where two-body quasiparticle scatterings are included.
Hence, adopting this approximation on the Schwinger-Keldysh contour,\cite{Kita10,HJ08,Rammer07} we can describe thermalization of superconductors, for example.

We now introduce the FLEX-S approximation, which adds the sort of diagrams (3) and (4a) in Fig.\ 5 up to infinite order
besides $\Phi_{1}$ and $\Phi_{2}$.
As seen from Fig.\ 1, it contains the normal particle-hole and particle-particle diagrams, i.e., diagrams (a) and (b), respectively, where every exchange process is automatically incorporated through $\Gamma^{(0)}$, plus all the pair processes derivable from them
by successively changing directions of a pair of incoming and outgoing arrows at a vertex, i.e., diagrams (c)-(j).
Generalizing the consideration that led to eqs.\ (\ref{Phi_3}) and (\ref{Phi_4a}), we find the $n$th-order contribution $\Phi_{na}$ ($n\geq 4$) as
\begin{equation}
\Phi_{na}=\frac{(-1)^{n+1}}{2n\beta}{\rm Tr}\,\bigl(\underline{\Gamma}^{(0)}\underline{\chi}^{(0)}\bigr)^{\! n}.
\end{equation}
The functional $\Phi$ of our FLEX-S approximation is defined by
$\Phi_{\mbox{FLEX-S}}=\Phi_{1}+\Phi_{2}+\Phi_{3}+\sum_{n=4}^{\infty}\Phi_{na}$,
i.e.,
\begin{eqnarray}
&&\hspace{-13mm}
\Phi_{\mbox{FLEX-S}}=\frac{1}{4\beta}{\rm Tr}\, \underline{\Gamma}^{(0)}\underline{\chi}^{(0)}+
\frac{1}{6\beta}{\rm Tr}\bigl(\underline{\Gamma}^{(0)}\underline{\chi}^{(0)}\bigr)^{2}
\nonumber \\
&&\hspace{7mm}
+\frac{1}{2\beta}{\rm Tr} \! \left[\ln \bigl(\underline{1}+\underline{\Gamma}^{(0)}\underline{\chi}^{(0)}\bigr)
-\underline{\Gamma}^{(0)}\underline{\chi}^{(0)}\right] .
\label{Phi_FLEX}
\end{eqnarray}
The corresponding self-energy is obtained by eq.\ (\ref{Sigma-Phi}) with $\Phi\approx \Phi_{\mbox{FLEX-S}}$,
where the differentiation may also be carried out graphically by removing a single line from the diagrams for $\Phi$ in all possible ways.
Using eq.\ (\ref{hatGamma-symm}), we arrive at the expression
\begin{eqnarray}
&&\hspace{-8mm}
\Sigma_{ij}(1,2)=-\biggl[\langle 1\bar{3}_{i\bar{k}}|\underline{\Gamma}^{(0)} |2\bar{4}_{j\bar{l}}\rangle
\!+\! \frac{4}{3}\langle 1\bar{3}_{i\bar{k}}|\underline{\Gamma}^{(0)}\underline{\chi}^{(0)}\underline{\Gamma}^{(0)}|2\bar{4}_{j\bar{l}}\rangle
\nonumber \\
&&\hspace{9.5mm}
-2\langle 1\bar{3}_{i\bar{k}}|\underline{\Gamma}^{(0)}\underline{\chi}^{(0)}\underline{\Gamma}^{(1)}|2\bar{4}_{j\bar{l}}\rangle 
\biggr] G_{\bar{k}\bar{l}}(\bar{3},\bar{4}) ,
\label{Sigma_FLEX}
\end{eqnarray}
where $\underline{\Gamma}^{(1)}$ is defined by
\begin{equation}
\underline{\Gamma}^{(1)}\equiv \underline{\Gamma}^{(0)}(\underline{1}+\underline{\chi}^{(0)}\underline{\Gamma}^{(0)})^{-1}.
\label{Gamma^(1)-def}
\end{equation}

\subsection{Two-particle Green's function}

Next, we consider the two-particle Green's function
\begin{eqnarray}
&&\hspace{-10mm}
{\cal K}_{ii',jj'}(11',22')
\equiv \langle T_{\tau}\psi_{i}^{\rm H}(1)\psi_{3-i'}^{\rm H}(1')\psi_{j}^{\rm H}(2)\psi_{3-j'}^{\rm H}(2')\rangle 
\nonumber \\
&&\hspace{18mm}
-G_{ii'}(1,1')G_{jj'}(2,2') ,
\label{calK-def}
\end{eqnarray}
whose poles define collective modes such as density, spin, and superconducting fluctuations.
Functional $\Phi[\hat{G}]$ also enables us to calculate eq.\ (\ref{calK-def})  unambiguously.
To see this, let us apply an artificial nonlocal potential $\hat{\cal U}(1,2)$ described by the S-matrix
\begin{equation}
{\cal S}^{\cal U}\equiv T_{\tau}
\exp\!\left[ \,\frac{1}{2} \psi_{\bar{j}}^{\rm H}(\bar{2})\psi_{3-\bar{j}'}^{\rm H}(\bar{2}') {\cal U}_{\bar{j}'\bar{j}}(\bar{2}',\bar{2})\right]\! .
\label{calS_ex}
\end{equation}
Green's function in the presence of $\hat{\cal U}\equiv ({\cal U}_{ij})$ is defined by\cite{AGD63}
\begin{equation}
G^{\cal U}_{ii'}(1,1')\equiv -\frac{\langle T_{\tau}{\cal S}^{\cal U}\psi_{i}^{\rm H}(1)\psi_{3-i'}^{\rm H}(1')\rangle}
{\langle{\cal S}^{\cal U}\rangle} ,
\label{G_ij-U}
\end{equation}
which reduces to eq.\ (\ref{G_ij}) as $\hat{\cal U}\rightarrow \hat{0}$. 
Now, one may see easily that eq.\ (\ref{calK-def}) 
is obtained from eq.\ (\ref{G_ij-U}) by
\begin{equation}
{\cal K}_{ii',jj'}(11',22')=-2\frac{\delta G^{\cal U}_{ii'}(1,1')}{\delta {\cal U}_{j'j}(2',2)}\biggr|_{\hat{\cal U}\rightarrow\hat{0}}.
\label{calK-deriv1}
\end{equation}
This expression tells us that we only need to know the linear response of $\hat{G}^{\cal U}$ to $\hat{\cal U}$ for obtaining the two-particle Green's function.

To find $\delta \hat{G}^{\cal U}$, we start from the Dyson-Gor'kov equation, which is modified from eq.\ (\ref{DG}) into\cite{MS59,BK61,Kita10b} 
\begin{equation}
\bigl[ \hat{G}_{0}^{-1}(1,\bar{2})-\hat{\cal U}'(1,\bar{2})-\hat{\Sigma}^{\cal U\!}(1,\bar{2})\bigr]\hat{G}^{\cal U\!}(\bar{2},1')=\hat{\sigma}_{0}\delta(1,1') ,
\label{DG-U}
\end{equation}
with
\begin{equation}
{\cal U}_{ii'}'(1,1')\equiv \frac{{\cal U}_{ii'}(1,1')-{\cal U}_{3-i',3-i}(1',1)}{2} .
\label{U'}
\end{equation}
Let us change $\hat{\cal U}\rightarrow \hat{\cal U}+\delta \hat{\cal U}$ in eq.\ (\ref{DG-U}) and subsequently set $\hat{\cal U}=\hat{0}$.
We thereby obtain the first-order equation
\begin{equation}
\hat{G}^{-1}(1,\bar{2})\delta \hat{G}^{\cal U\!}(\bar{2},1')
=\bigl[\delta \hat{\cal U}'(1,\bar{2}) +\delta \hat{\Sigma}^{\cal U\!}(1,\bar{2})\bigr]\hat{G}(\bar{2},1').
\label{DG-dU}
\end{equation}
Using eq.\ (\ref{Sigma-Phi}) for the self-energy, we can express $\delta \hat{\Sigma}^{\cal U}$ above in terms of 
$\delta \hat{G}^{\cal U}$ as
\begin{equation}
\delta \Sigma_{ii'}^{\cal U}(1,1')=\Gamma^{({\rm ir})}_{ii',\bar{j}\bar{j}'}(11',\bar{2}\bar{2}')\delta G_{\bar{j}'\bar{j}}^{\cal U}(\bar{2}',\bar{2}) ,
\label{dSigma}
\end{equation}
where $\Gamma^{({\rm ir})}_{ii',\bar{j}\bar{j}'}(11',\bar{2}\bar{2}')$ denotes the ``irreducible'' vertex defined by
\begin{equation}
\Gamma^{({\rm ir})}_{ii',\bar{j}\bar{j}'}(11',22')\equiv 2\beta \frac{\delta^{2}\Phi }{\delta G_{i'i}(1',1)\delta G_{j'j}(2',2)}.
\label{Gamma^(ir)}
\end{equation}
This differentiation should be performed after the replacement of eq.\ (\ref{tG_ij}) in $\Phi$, as already mentioned.
Let us substitute eq.\ (\ref{dSigma}) into eq.\ (\ref{DG-dU}), multiply the resultant equation by $\hat{G}(3,1)$ from the left, and integrate over $1$.
Changing the arguments appropriately, we obtain a closed equation for $\delta\hat{G}^{\cal U}$ as 
\begin{eqnarray}
&&\hspace{-9mm}
\delta G^{\cal U}_{ii'}(1,1')
= G_{i\bar{j}}(1,\bar{2})G_{\bar{j}'i'}(\bar{2}',1')\delta {\cal U}_{\bar{j}\bar{j}'}'(\bar{2},\bar{2}')
+ G_{i\bar{j}}(1,\bar{2})
\nonumber \\
&&\hspace{12mm}
\times  G_{\bar{j'}i'}(\bar{2}',1')
\Gamma^{({\rm ir})}_{\bar{j}\bar{j}',\bar{k}'\bar{k}}(\bar{2}\bar{2}',\bar{3}'\bar{3})\delta G^{\cal U}_{\bar{k}\bar{k}'}(\bar{3},\bar{3}').
\nonumber \\
\label{dG-eq}
\end{eqnarray}

At this stage, it is convenient to introduce the notations
\begin{equation}
\langle 11'_{ii'}|\delta\vec{G}^{\cal U}=\delta G_{ii'}^{\cal U}(1,1'),
\hspace{5mm}
\langle 11'_{ii'}|\delta\vec{\cal U}=\delta {\cal U}_{ii'}(1,1') ,
\label{Vec3}
\end{equation}
and 
\begin{subequations}
\label{Mat}
\begin{eqnarray}
&&\hspace{-8mm}
\langle 11'_{ii'}|\underline{\cal K}|22'_{jj'}\rangle\equiv
{\cal K}_{ii',j'j}(11',2'2),
\label{calK-mat}
\\
&&\hspace{-11.5mm}
\langle 11'_{ii'}|\underline{\Gamma}^{({\rm ir})}|22'_{jj'}\rangle\equiv
\Gamma^{({\rm ir})}_{ii',j'j}(11',2'2),
\label{Gamma^(ir)-mat}
\\
&&\hspace{-12mm}
\langle 11'_{ii'}|\underline{\chi}^{(0{\rm e})}|22'_{jj'}\rangle\equiv -G_{ij}(1,2)G_{j'i'}(2',1')
\nonumber \\
&&\hspace{18.5mm} +G_{i,3-j'}(1,2')G_{3-j,i'}(2,1'),
\label{t-chi^(0)}
\\
&&\hspace{-6.5mm}
\langle 11'_{ii'}|\underline{1}|22'_{jj'}\rangle\equiv \delta_{ij}\delta_{i'j'}\delta(1,2)\delta(1',2').
\label{u1}
\end{eqnarray}
\end{subequations}
It follows from eqs.\ (\ref{G_ij-symm}), (\ref{calK-def}), (\ref{Gamma^(ir)}), and (\ref{Mat}) that matrices $\underline{\cal M}= \underline{\cal K}$,  
$\underline{\Gamma}^{({\rm ir})}$, $\underline{\chi}^{(0{\rm e})}$ all satisfy
\begin{eqnarray}
&&\hspace{-10mm}
\langle 11'_{ii'}|\underline{\cal M}|22'_{jj'}\rangle=\langle 2'2_{j'j}|\underline{\cal M}|1'1_{i'i}\rangle
\nonumber \\
&&\hspace{14mm}
=-\langle 1'1_{3-i',3-i}|\underline{\cal M}|22'_{jj'}\rangle .
\label{GC-symm}
\end{eqnarray}
The first equality is also obeyed by $\underline{\chi}^{(0)}$ of eq.\ (\ref{chi^(0)}).

\begin{figure}[t]
        \begin{center}
                \includegraphics[height=10mm]{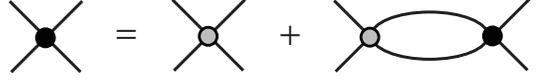}
        \end{center}
        \caption{Bethe-Salpeter equation for superconductivity. A black (shaded) circle denotes $\underline{\Gamma}$ ($\underline{\Gamma}^{({\rm ir})}$), and 
        an internal line represents $\hat{G}$. \label{Fig6}}
\end{figure}

Using eqs.\ (\ref{chi^(0)}), (\ref{U'}), (\ref{Vec3}), and (\ref{Mat}), we can express eq.\ (\ref{dG-eq}) as
$\delta \vec{G}^{\cal U}= -\frac{1}{2}\underline{\chi}^{(0{\rm e})}\delta\vec{\cal U}
-\underline{\chi}^{(0)}\underline{\Gamma}^{({\rm ir})}\delta \vec{G}^{\cal U}$, or equivalently, 
$$\delta\vec{G}^{\cal U} =- \frac{1}{2}\bigl(\underline{1}+\underline{\chi}^{(0)}\underline{\Gamma}^{({\rm ir})}\bigr)^{-1}\underline{\chi}^{(0{\rm e})} \delta\vec{\cal U}.$$
Finally, we obtain an expression of $\underline{\cal K}$ by eq.\ (\ref{calK-deriv1}) as
\begin{equation}
\underline{\cal K} = \bigl(\underline{1}+\underline{\chi}^{(0)}\underline{\Gamma}^{({\rm ir})}\bigr)^{-1}\underline{\chi}^{(0{\rm e})}  .
\label{calK}
\end{equation}
It may be useful to introduce the full vertex
\begin{equation}
\underline{\Gamma} \equiv \underline{\Gamma}^{({\rm ir})}\bigl(\underline{1}+\underline{\chi}^{(0)}\underline{\Gamma}^{({\rm ir})}\bigr)^{-1} ,
\label{BS}
\end{equation}
which is nothing but the Bethe-Salpeter equation\cite{SB51} for superconductivity and given graphically by Fig.\ \ref{Fig6}. Using 
$\underline{\Gamma}$, we can express eq.\ (\ref{calK}) alternatively as
\begin{equation}
\underline{\cal K} = \underline{\chi}^{(0{\rm e})} -\underline{\chi}^{(0)}\underline{\Gamma}\underline{\chi}^{(0{\rm e})} .
\label{calK2}
\end{equation}
Equation (\ref{calK}) or (\ref{calK2}) with eqs.\ (\ref{chi^(0)}), (\ref{Gamma^(ir)}), and (\ref{Mat}) tells us that we can also calculate the two-particle Green's function of 
eq.\ (\ref{calK-def}) once $\Phi$ is given explicitly. 

It may be illuminating to write down $\underline{\Gamma}^{({\rm ir})}$ explicitly in the FLEX-S approximation.
Let us make the replacement of eq.\ (\ref{tG_ij}) in eq.\ (\ref{Phi_FLEX}), substitute the resultant $\Phi$ into eq.\ (\ref{Gamma^(ir)}),
and perform the differentiation.
Also using the symmetry noted around eq.\ (\ref{GC-symm}), we obtain
\begin{eqnarray}
&&\hspace{-10mm}
 \langle 11'_{ii'}|\underline{\Gamma}^{({\rm ir})}|22'_{jj'}\rangle
\nonumber \\
&&\hspace{-14mm}
= \langle 11'_{ii'}|\underline{\Gamma}^{(0)}|22'_{jj'}\rangle 
-2\bigl[ \langle 12_{ij}|\underline{\Gamma}^{(0)}\underline{\chi}^{(0)}\underline{\Gamma}^{(0)}|1'2'_{i'j'}\rangle 
\nonumber \\
&&\hspace{-10mm}
-\langle 12'_{i,3-j'}|\underline{\Gamma}^{(0)}\underline{\chi}^{(0)}\underline{\Gamma}^{(0)}|1'2_{i',3-j}\rangle \bigr]
\nonumber \\
&&\hspace{-10mm}
+\langle 12_{ij}|\underline{\Gamma}^{(0)}\underline{\chi}^{(0)}\underline{\Gamma}^{(1)}|1'2'_{i'j'}\rangle 
\nonumber \\
&&\hspace{-10mm}
-\langle 12'_{i,3-j'}|\underline{\Gamma}^{(0)}\underline{\chi}^{(0)}\underline{\Gamma}^{(1)}|1'2_{i',3-j}\rangle 
\nonumber \\
&&\hspace{-10mm}
+2\bigl[\langle 1\bar{3}_{i\bar{k}}|\underline{\Gamma}^{(1)}|2\bar{4}'_{{j}\bar{l}'}\rangle
\langle2'\bar{4}_{j'\bar{l}}|\underline{\Gamma}^{(1)}|1'\bar{3}'_{i'\bar{k}'}\rangle 
\nonumber \\
&&\hspace{-10mm}
-\langle 1\bar{3}_{i\bar{k}}|\underline{\Gamma}^{(1)}|2'\bar{4}'_{3-{j}',\bar{l}'}\rangle
\langle2\bar{4}_{3-j,\bar{l}}|\underline{\Gamma}^{(1)}|1'\bar{3}'_{i'\bar{k}'}\rangle  \bigr]
\nonumber \\
&&\hspace{-10mm}
\times
\langle\bar{3}\bar{4}'_{\bar{k}\bar{l}'}|\underline{\chi}^{(0)}|\bar{3}'\bar{4}_{\bar{k}'\bar{l}}\rangle ,
\end{eqnarray}
which is different from $\underline{\Gamma}^{(1)}$ of eq.\ (\ref{Gamma^(1)-def}). 
Thus, a clear distinction among $\underline{\Gamma}^{(1)}$, $\underline{\Gamma}^{({\rm ir})}$, and $\underline{\Gamma}$
is necessary.

\subsection{Alternative approach}

Before closing this section, we comment on an alternative self-consistent approach adopted by Haussmann {\em et al}.\ \cite{Haussmann07} 
of expressing $\Omega$ in terms of a renormalized vertex $\Gamma^{\Omega}$ besides $\hat{G}$. 
Its theoretical basis was given by de Dominicis and Martin;\ \cite{dDM64}
see also refs.\ \onlinecite{deDominicis63} and \onlinecite{Bloch65} on this point.
However, the vertices $\Gamma^{\Omega}$ obtained by de Dominicis \cite{deDominicis63,Bloch65} and used by Haussmann {\em et al}.\ \cite{Haussmann07} 
have no internal degrees of freedom describing only density fluctuations, in contrast to our $\underline{\Gamma}^{(0)}$ of eq.\ (\ref{hatGamma^(0)}).
Indeed, constructing $\Omega=\Omega[\hat{G}, \Gamma^{\Omega}]$ with a matrix structure in $\Gamma^{\Omega}$ seems to be a non-trivial problem.
It will be practically more convenient to use $\Omega=\Omega[\hat{G}]$ that can easily be written down solely in terms of the smallest expansion unit $\hat{G}$.
Note in this context that: (i) the results obtained by using $\Omega=\Omega[\hat{G}, \Gamma^{\Omega}]$ should necessarily be reproduced by the approach based
on $\Omega=\Omega[\hat{G}]$, as in the case of the particle-particle ladder approximation by Haussmann {\em et al}.; \cite{Haussmann07}
(ii) the vertex $\Gamma^{\Omega}$ in $\Omega$ is generally different from both the irreducible vertex $\Gamma^{({\rm ir})}$ of eq.\ (\ref{Gamma^(ir)}) and the full vertex $\Gamma$
of eq.\ (\ref{BS}).

\section{Homogeneous systems}

We now simplify the formalism of \S 2 for homogeneous systems.
Lattice models may be handled similarly with modifications given in Appendix\ref{appendix:Hubbard}.

\subsection{Momentum-``energy'' representation}

Equations (\ref{G_ij}) and (\ref{hatSigma}) and the delta function $\delta(1,2)$ can be expanded for homogeneous systems as
\begin{subequations}
\label{GS-exp}
\begin{eqnarray}
&&\hspace{-10mm}
G_{ij}(1,2)=\sum_{\vec{p}}G_{i\alpha,j\beta}(\vec{p})\,{\rm e}^{i\vec{p}\cdot(\vec{r}_{1}-\vec{r}_{2})},
\label{G_ij-exp}
\\
&&\hspace{-9.5mm}
\Sigma_{ij}(1,2)=\sum_{\vec{p}}\Sigma_{i\alpha,j\beta}(\vec{p})\,{\rm e}^{i\vec{p}\cdot(\vec{r}_{1}-\vec{r}_{2})},
\label{Sigma_ij-exp}
\\
&&\hspace{-6.2mm}
\delta(1,2)=\delta_{\alpha\beta}\sum_{\vec{p}}{\rm e}^{i\vec{p}\cdot(\vec{r}_{1}-\vec{r}_{2})}.
\label{delta-fn}
\end{eqnarray}
\end{subequations}
Here $\vec{r}_{1}\!\equiv\! ({\bm r}_{1},i\tau_{1})$ and $\vec{p}\!\equiv\! ({\bm p},i\varepsilon_{n})$ with $\varepsilon_{n}\!\equiv\! (2n+1)\pi T$ the fermion Matsubara frequency ($n=0,\pm 1,\cdots$), subscripts $_\alpha, _\beta$ denote spin indices, and the summation over $\vec{p}$ is defined by
\begin{equation}
\sum_{\vec{p}}\equiv \frac{1}{\beta}\sum_{n}\int \frac{{\rm d}^{3}p}{(2\pi)^{3}}.
\label{p-sum}
\end{equation}
We call the imaginary quantity $i\varepsilon_{n}$ ``energy'' below.
Substituting eq.\ (\ref{GS-exp}) into eq.\ (\ref{DG}),
we can transform the Dyson-Gor'kov equation into the $4\times 4$ matrix equation
\begin{equation}
\bigl[\hat{G}_{0}^{-1}(\vec{p}\,)-\hat{\Sigma}(\vec{p}\,)\bigr]\hat{G}(\vec{p}\,)=\hat{1},
\label{DG(p)}
\end{equation}
where $\hat{G}_{0}^{-1}(\vec{p}\,)$ and $\hat{1}$ are defined by
\begin{equation}
\langle i\alpha|\hat{G}_{0}^{-1}(\vec{p}\,)|j\beta\rangle=\delta_{ij}\delta_{\alpha\beta}\bigl[i\varepsilon_{n}-(-1)^{i-1}\epsilon_{\bm p}\bigr]
\end{equation}
with $\epsilon_{\bm p}\equiv p^{2}/2m-\mu$ and $\langle i\alpha|\hat{1}|j\beta\rangle=\delta_{ij}\delta_{\alpha\beta}$, respectively.
The standard arrangement of the basis vectors for the direct product ${\rm C}\times {\rm S}$ of the particle-hole (i.e., charge) space C ($i=1,2$) and spin space S ($\alpha=\uparrow,\downarrow$) is given by $|{\rm cs}_{1}\rangle=|1\!\uparrow\rangle$, $|{\rm cs}_{2}\rangle=|1\!\downarrow\rangle$,  $|{\rm cs}_{3}\rangle=|2\!\uparrow\rangle$, and  $|{\rm cs}_{4}\rangle=|2\!\downarrow\rangle$. 
Green's function in this representation reads
\begin{equation}
\hat{G}(\vec{p})=\begin{bmatrix}
\vspace{1mm}
\underline{G}(\vec{p}) & \underline{F}(\vec{p}) \\
- \underline{\bar{F}}(\vec{p}) &  -\underline{\bar{G}}(\vec{p})
\end{bmatrix},
\label{hatG(p)}
\end{equation}
where each quantity with an underline is a $2\times 2$ matrix in spin space given as a Fourier coefficient of eq.\ (\ref{GF-def}).
It follows from eqs.\ (\ref{G_ij-symm}) and (\ref{GF-def}) that these submatrices satisfy
\begin{subequations}
\label{GF-symm}
\begin{equation}
G_{\alpha\beta}(\vec{p})=\bar{G}_{\beta\alpha}(-\vec{p})=G_{\beta\alpha}^{*}(\vec{p}^{\,*}),
\end{equation}
\begin{equation}
F_{\alpha\beta}(\vec{p})=-F_{\beta\alpha}(-\vec{p})=-\bar{F}_{\beta\alpha}^{*}(\vec{p}^{\,*}).
\end{equation}
\end{subequations}
The self-energy matrix may also be expressed as
\begin{equation}
\hat{\Sigma}(\vec{p})=\begin{bmatrix}
\vspace{1mm}
\underline{\Sigma}(\vec{p}) & \underline{\Delta}(\vec{p}) \\
- \underline{\bar{\Delta}}(\vec{p}) &  -\underline{\bar{\Sigma}}(\vec{p})
\end{bmatrix},
\label{hatSigma(p)}
\end{equation}
whose submatrices clearly obey the relations of eq.\ (\ref{GF-symm}).

Next, we transform $\hat{\Sigma}=\hat{\Sigma}[\hat{G}]$ into the momentum-energy representation.
Since the expression depends on the approximation we adopt,
we specifically consider the FLEX-S approximation given by eq.\ (\ref{Sigma_FLEX}). 
To begin with, we expand $V({\bm r}_{1}-{\bm r}_{2})\delta(\tau_{1}-\tau_{2})$ in eq.\ (\ref{Gamma^(0)}) as
\begin{equation}
V({\bm r}_{1}-{\bm r}_{2})\delta(\tau_{1}-\tau_{2})=\sum_{\vec{q}} V_{\bm q}\,{\rm e}^{i\vec{q}\cdot(\vec{r}_{1}-\vec{r}_{2})},
\label{V-exp}
\end{equation}
where $\vec{q}\equiv ({\bm q},i\omega_{\ell})$ with  $\omega_{\ell}\equiv 2\ell \pi T$ the boson Matsubara frequency ($\ell=0,\pm 1, \cdots$),
and the summation over $\vec{q}$ is defined in the same way as eq.\ (\ref{p-sum}) with the replacement $p\rightarrow q$ and $n\rightarrow \ell$.
Let us substitute eq.\ (\ref{Gamma^(0)}) into eq.\  (\ref{hatGamma^(0)-2}), use eqs.\ (\ref{delta-fn}) and (\ref{V-exp}) subsequently, and make some changes of integration variables; see also Appendix\ref{appendix:Hubbard} for details. We thereby obtain an expansion of eq.\ (\ref{uGamma^(0)}) as
\begin{eqnarray}
&&\hspace{-10mm}
\langle 11'_{ii'}|\underline{\Gamma}^{(0)}|22'_{jj'}\rangle
= \sum_{\vec{p}\vec{p}^{\,\prime}\vec{q}}
\langle i\alpha, i'\alpha'|\underline{\Gamma}^{(0)}({\bm p},{\bm p}',{\bm q}) | j\beta, j'\beta'\rangle
\nonumber \\
&&\hspace{19mm}\times
{\rm e}^{i\vec{p}\cdot\vec{r}_{1}
-i\vec{p}_{-}\cdot\vec{r}_{1}^{\,\prime}
-i\vec{p}^{\,\prime}\cdot\vec{r}_{2}+i\vec{p}^{\,\prime}_{-}\cdot\vec{r}_{2}^{\,\prime}},
\label{hatGamma^(0)-exp}
\end{eqnarray}
where $\vec{p}_{-}$ is defined by
\begin{equation}
\vec{p}_{-}\equiv ({\bm p}-{\bm q},i\varepsilon_{n}-i\omega_{\ell}),
\end{equation}
and the Fourier coefficient is given by
\begin{eqnarray}
&&\hspace{-6mm}
\langle i\alpha, i'\alpha'|\underline{\Gamma}^{(0)}({\bm p},{\bm p}^{\prime},{\bm q}) | j\beta, j'\beta'\rangle
\nonumber \\
&&\hspace{-10mm}
=\frac{1}{2}\bigl[
(-1)^{i+j}\delta_{ii'}\delta_{jj'}\delta_{\alpha'\alpha}\delta_{\beta'\beta}V_{{\bm q}}
\nonumber \\
&&\hspace{-6mm}
-(-1)^{i+i'}\delta_{ij}\delta_{i'j'}\delta_{\alpha\beta}\delta_{\alpha'\beta'}V_{{\bm p}-{\bm p}'}
\nonumber \\
&&\hspace{-6mm}
+(-1)^{i+i'}\delta_{j',3-i}\delta_{j,3-i'}\delta_{\alpha\beta'}\delta_{\alpha'\beta}
V_{{\bm p}+{\bm p}'-{\bm q}}\bigr] .
\label{Gamma-p}
\end{eqnarray}
Note that the arguments $\tau_{1}$, $\tau_{1}'$, $\tau_{2}$, and $\tau_{2}'$ in eq.\ (\ref{hatGamma^(0)-exp}) are properly expanded 
in fermion Matsubara frequencies;
this is why we have expressed eq.\ (\ref{V-exp}) in terms of boson Matsubara frequencies.
Using eq.\ (\ref{G_ij-exp}), we can also transform eq.\ (\ref{chi^(0)}) into
\begin{eqnarray}
&&\hspace{-10mm}
\langle 11'_{ii'}|\underline{\chi}^{(0)}|22'_{jj'}\rangle
= \sum_{\vec{p}\vec{p}^{\,\prime}\vec{q}}
\langle i\alpha, i'\alpha'|\underline{\chi}^{(0)}(\vec{p},\vec{q}) | j\beta, j'\beta'\rangle\delta_{\vec{p}\vec{p}^{\,\prime}}
\nonumber \\
&&\hspace{20mm}\times
{\rm e}^{i\vec{p}\cdot\vec{r}_{1}
-i\vec{p}_{-}\cdot\vec{r}_{1}^{\,\prime}
-i\vec{p}^{\,\prime}\cdot\vec{r}_{2}+i\vec{p}^{\,\prime}_{-}\cdot\vec{r}_{2}^{\,\prime}},
\label{hatChi^(0)-exp}
\end{eqnarray}
with
\begin{equation}
\langle i\alpha, i'\alpha'|\underline{\chi}^{(0)}(\vec{p},\vec{q}) | j\beta, j'\beta'\rangle
=-G_{i\alpha,j\beta}(\vec{p}\,)G_{j'\beta',i'\alpha'}(\vec{p}_{-}).
\label{chi^(0)(p,q)}
\end{equation}
Substituting eqs.\ (\ref{GS-exp}), (\ref{hatGamma^(0)-exp}), and (\ref{hatChi^(0)-exp}) into eq.\ (\ref{Sigma_FLEX}),
we obtain the self-energy for homogeneous systems in the FLEX-S approximation as
\begin{eqnarray}
&&\hspace{-9mm}
\Sigma_{i\alpha,j\beta}(\vec{p})
=\sum_{\vec{q}}{\rm Tr}\biggl[\underline{\Gamma}^{(0)}({\bm p},{\bm p},{\bm q})
\nonumber \\
&&\hspace{9mm}
+\frac{4}{3}\sum_{\vec{p}_{1}}\underline{\Gamma}^{(0)}({\bm p},{\bm p}_{1},{\bm q}) \underline{\chi}^{(0)}(\vec{p}_{1},\vec{q})
\underline{\Gamma}^{(0)}({\bm p}_{1},{\bm p},{\bm q})
\nonumber \\
&&\hspace{9mm}
-2\sum_{\vec{p}_{1}}\underline{\Gamma}^{(0)}({\bm p},{\bm p}_{1},{\bm q}) \underline{\chi}^{(0)}(\vec{p}_{1},\vec{q})
\underline{\Gamma}^{(1)}({\bm p}_{1},{\bm p},\vec{q})\biggr]
\nonumber \\
&&\hspace{9mm}
\times \underline{\delta\chi}^{(i\alpha,j\beta)}(\vec{p}-\vec{q}),
\label{Sigma_FLEX(p)}
\end{eqnarray}
where $\underline{\Gamma}^{(1)}$ is the momentum-energy representation of eq.\ (\ref{Gamma^(1)-def}) satisfying
\begin{eqnarray}
&&\hspace{-10mm}
\underline{\Gamma}^{(1)}({\bm p},{\bm p}',\vec{q})=\underline{\Gamma}^{(0)}({\bm p},{\bm p}',{\bm q})-
\sum_{\vec{p}_{1}}\underline{\Gamma}^{(0)}({\bm p},{\bm p}_{1},{\bm q})
\nonumber \\
&&\hspace{13mm}
\times \underline{\chi}^{(0)}(\vec{p}_{1},\vec{q})
\underline{\Gamma}^{(1)}({\bm p}_{1},{\bm p}',\vec{q}),
\label{Gamma^(1)(p)}
\end{eqnarray}
and $\underline{\delta\chi}^{(i\alpha,j\beta)}(\vec{q})$ is defined by
\begin{eqnarray}
&&\hspace{-6mm}
\langle j_{1}\beta_{1},j_{2}\beta_{2}| \underline{\delta\chi}^{(i\alpha,j\beta)}(\vec{p})|  i_{1}\alpha_{1},i_{2}\alpha_{2}\rangle
\nonumber \\
&&\hspace{-10mm}
\equiv -\delta_{j_{1}j}\delta_{\beta_{1}\beta}\delta_{i_{1}i}\delta_{\alpha_{1}\alpha} G_{i_{2}\alpha_{2},j_{2}\beta_{2}}(\vec{p}) .
\label{dchi-def}
\end{eqnarray}

The basis vectors of Tr in eq.\ (\ref{Sigma_FLEX(p)}) are given by $|i\alpha,i'\alpha'\rangle$, 
which belong to the direct product $({\rm C}\times {\rm S})\times ({\rm C}\times {\rm S})$
with dimension $2^{4}$. 
It is convenient to arrange them as $|({\rm cs})^{2}_{\nu}\rangle$ $(\nu=1,\cdots,16)$ with
\begin{eqnarray}
\begin{array}{ll}
 |({\rm cs})^{2}_{1}\rangle =|1\uparrow,1\uparrow\rangle , \,\,\,&\,\,\,
|({\rm cs})^{2}_{2}\rangle =|1\uparrow,1\downarrow\rangle ,\\
|({\rm cs})^{2}_{3}\rangle =|1\downarrow,1\uparrow\rangle , \,\,\,&\,\,\,
|({\rm cs})^{2}_{4}\rangle =|1\downarrow,1\downarrow\rangle ,\\
|({\rm cs})^{2}_{5}\rangle =|2\uparrow,2\uparrow\rangle , \,\,\,&\,\,\,
|({\rm cs})^{2}_{6}\rangle =|2\uparrow,2\downarrow\rangle ,\\
|({\rm cs})^{2}_{7}\rangle =|2\downarrow,2\uparrow\rangle , \,\,\,&\,\,\,
|({\rm cs})^{2}_{8}\rangle =|2\downarrow,2\downarrow\rangle ,\\
|({\rm cs})^{2}_{9}\rangle =|1\uparrow,2\uparrow\rangle , \,\,\,&\,\,\,
|({\rm cs})^{2}_{10}\rangle =|1\uparrow,2\downarrow\rangle ,\\
|({\rm cs})^{2}_{11}\rangle =|1\downarrow,2\uparrow\rangle , \,\,\,&\,\,\,
|({\rm cs})^{2}_{12}\rangle =|1\downarrow,2\downarrow\rangle ,\\
|({\rm cs})^{2}_{13}\rangle =|2\uparrow,1\uparrow\rangle , \,\,\,&\,\,\,
|({\rm cs})^{2}_{14}\rangle =|2\uparrow,1\downarrow\rangle ,\\
|({\rm cs})^{2}_{15}\rangle =|2\downarrow,1\uparrow\rangle , \,\,\,&\,\,\,
|({\rm cs})^{2}_{16}\rangle =|2\downarrow,1\downarrow\rangle .
\end{array}
\label{BV2}
\end{eqnarray}
Then, $\underline{\chi}^{(0)}(\vec{p},\vec{q})$ in eq.\ (\ref{chi^(0)(p,q)}) can be represented by a $16\times 16$ matrix as
\begin{equation}
\underline{\chi}^{(0)}=
\begin{bmatrix}
-\underline{GG} & \underline{F\bar{F}} & \underline{G\bar{F}} & -\underline{FG} \\
\underline{\bar{F}F} & -\underline{\bar{G}\bar{G}} & -\underline{\bar{F}\bar{G}} & \underline{\bar{G}F} \\
-\underline{GF} & \underline{F\bar{G}} & \underline{G\bar{G}} & -\underline{FF} \\
\underline{\bar{F}G} & -\underline{\bar{G}\bar{F}} & -\underline{\bar{F}\bar{F}} & \underline{\bar{G}G}
\end{bmatrix},
\label{chi-mat}
\end{equation}
where each element is a $4\times 4$ submatrix given in terms of the elements of eq.\ (\ref{hatG(p)});
for example, $\underline{F}\underline{\bar{F}}$ denotes
\begin{equation}
\underline{F}\underline{\bar{F}}=
\begin{bmatrix}
F_{\uparrow\uparrow}\bar{F}_{\uparrow\uparrow} & F_{\uparrow\uparrow}\bar{F}_{\downarrow\uparrow}
& F_{\uparrow\downarrow}\bar{F}_{\uparrow\uparrow} & F_{\uparrow\downarrow}\bar{F}_{\downarrow\uparrow} \\
F_{\uparrow\uparrow}\bar{F}_{\uparrow\downarrow} & F_{\uparrow\uparrow}\bar{F}_{\downarrow\downarrow}
& F_{\uparrow\downarrow}\bar{F}_{\uparrow\downarrow} & F_{\uparrow\downarrow}\bar{F}_{\downarrow\downarrow} \\
F_{\downarrow\uparrow}\bar{F}_{\uparrow\uparrow} & F_{\downarrow\uparrow}\bar{F}_{\downarrow\uparrow}
& F_{\downarrow\downarrow}\bar{F}_{\uparrow\uparrow} & F_{\downarrow\downarrow}\bar{F}_{\downarrow\uparrow} \\
F_{\downarrow\uparrow}\bar{F}_{\uparrow\downarrow} & F_{\downarrow\uparrow}\bar{F}_{\downarrow\downarrow}
& F_{\downarrow\downarrow}\bar{F}_{\uparrow\downarrow} & F_{\downarrow\downarrow}\bar{F}_{\downarrow\downarrow} 
\end{bmatrix},
\label{FF-def}
\end{equation}
with $F_{\alpha\beta}\bar{F}_{\gamma\delta}=F_{\alpha\beta}(\vec{p})\bar{F}_{\gamma\delta}(\vec{p}_{-})$.
We can also express $\underline{\Gamma}^{(0)}({\bm p},{\bm p}',{\bm q})$ in eq.\ (\ref{hatGamma^(0)-exp}) as
\begin{equation}
\underline{\Gamma}^{(0)}=
\begin{bmatrix}
\underline{\Gamma}^{(0a)} & \underline{\Gamma}^{(0b)} &  \underline{0} & \underline{0} \\
\underline{\Gamma}^{(0b)} & \underline{\Gamma}^{(0a)} &  \underline{0} & \underline{0} \\
\underline{0} & \underline{0} & \underline{\Gamma}^{(0c)} & \underline{0} \\
\underline{0} & \underline{0} & \underline{0} & \underline{\Gamma}^{(0c)} 
\end{bmatrix} ,
\label{Gamma-mat}
\end{equation}
where $\underline{\Gamma}^{(0a,0b,0c)}$ are given by
\begin{subequations}
\label{Gamma-mat^(a,b,c)}
\begin{eqnarray}
&&\hspace{-11mm}
\underline{\Gamma}^{(0a)}({\bm p},{\bm p}',{\bm q})
\equiv
\frac{V_{\bm q}}{2}\underline{\gamma}^a-\frac{V_{{\bm p}-{\bm p}'}}{2}\underline{1},
\\
&&\hspace{-10.4mm}
\underline{\Gamma}^{(0b)}({\bm p},{\bm p}',{\bm q})
\equiv
\frac{V_{{\bm p}+{\bm p}'-{\bm q}}}{2}\underline{\gamma}^b-\frac{V_{\bm q}}{2}\underline{\gamma}^a,
\\
&&\hspace{-10mm}
\underline{\Gamma}^{(0c)}({\bm p},{\bm p}',{\bm q})
\equiv
\frac{V_{{\bm p}-{\bm p}'}}{2}\underline{1}-\frac{V_{{\bm p}+{\bm p}'-{\bm q}}}{2}\underline{\gamma}^b,
\end{eqnarray}
\end{subequations}
with $\underline{1}$ denoting the $4\times 4$ unit matrix and 
\begin{equation}
\underline{\gamma}^{a}\equiv 
\begin{bmatrix}
1 & 0 & 0 & 1 \\
0 & 0 & 0 & 0 \\
0 & 0 & 0 & 0 \\
1 & 0 & 0 & 1
\end{bmatrix},\hspace{5mm}
\underline{\gamma}^{b}\equiv
\begin{bmatrix}
1 & 0 & 0 & 0 \\
0 & 0 & 1 & 0 \\
0 & 1 & 0 & 0 \\
0 & 0 & 0 & 1
\end{bmatrix}.
\end{equation}
Equation (\ref{dchi-def}) may also be transformed into a matrix representation.

Equations (\ref{DG(p)}) and (\ref{Sigma_FLEX(p)}) form a closed set of self-consistent equations
for $\hat{G}(\vec{p})$ in the FLEX-S approximation, which is applicable to an arbitrary pairing symmetry 
including a mixture of singlet and triplet pairings.
By using the representation of eq.\ (\ref{BV2}), practical calculations may be performed based on the standard 
matrix algebra plus treatment of integral equations.
Some unitary transformations will be helpful for simplifying those calculations. 
For example, consider the orthogonal matrix
\begin{equation}
\underline{R}=
\begin{bmatrix}
\underline{R}^{a} & -\underline{R}^{b} &  \underline{0} & \underline{0} \\
\underline{R}^{b} & \underline{R}^{a} &  \underline{0} & \underline{0} \\
\underline{0} & \underline{0} & \underline{R}^{c} & \underline{0} \\
\underline{0} & \underline{0} & \underline{0} & \underline{R}^{c} 
\end{bmatrix} ,
\label{R-def}
\end{equation}
with
\begin{subequations}
\begin{equation}
\underline{R}^{a}\equiv 
\begin{bmatrix}
\frac{1}{2} & 0 & 0 & \frac{1}{2} \\
0 & \frac{1}{\sqrt{2}} & 0 & 0 \\
0 & 0 & \frac{1}{\sqrt{2}} & 0 \\
-\frac{1}{2} & 0 & 0 & \frac{1}{2}
\end{bmatrix},
\end{equation}
\begin{equation}
\underline{R}^{b}\equiv
\begin{bmatrix}
\frac{1}{2} & 0 & 0 & \frac{1}{2} \\
0  & 0 & \frac{1}{\sqrt{2}} & 0 \\
0 & \frac{1}{\sqrt{2}} & 0 & 0 \\
-\frac{1}{2} & 0 & 0 & \frac{1}{2}
\end{bmatrix},
\end{equation}
\begin{equation}
\underline{R}^{c}\equiv
\begin{bmatrix}
1 & 0 & 0 & 0 \\
0  & \frac{1}{\sqrt{2}} & -\frac{1}{\sqrt{2}} & 0 \\
0 & \frac{1}{\sqrt{2}} & \frac{1}{\sqrt{2}} & 0 \\
0 & 0 & 0 & 1
\end{bmatrix} .
\end{equation}
\end{subequations}
By using $\underline{R}$, eq.\ (\ref{Gamma-mat}) is transformed into the diagonal form
\begin{equation}
\underline{\tilde{\Gamma}}^{(0)}\equiv \underline{R}\,\underline{\Gamma}^{(0)}\underline{R}^{-1}=
\begin{bmatrix}
\underline{\tilde{\Gamma}}^{(0a)} & \underline{0} &  \underline{0} & \underline{0} \\
\underline{0} & \underline{\tilde{\Gamma}}^{(0b)} &  \underline{0} & \underline{0} \\
\underline{0} & \underline{0} & \underline{\tilde{\Gamma}}^{(0c)} & \underline{0} \\
\underline{0} & \underline{0} & \underline{0} & \underline{\tilde{\Gamma}}^{(0c)} 
\end{bmatrix} ,
\label{tGamma-mat}
\end{equation}
where $\underline{\tilde{\Gamma}}^{(0a,0b,0c)}({\bm p},{\bm p}',{\bm q})$ are given by
\begin{subequations}
\label{tGamma-mat^(a,b,c)}
\begin{eqnarray}
&&\hspace{-10.5mm}
\underline{\tilde{\Gamma}}^{(0a)}=
\begin{bmatrix}
V^{a} \!&\! 0 \!&\! 0 \!&\! 0 \\
0 \!&\! -V^{c} \!&\! 0 \!&\! 0 \\
0 \!&\! 0 \!&\! -V^{c} \!&\! 0 \\
0 \!&\! 0 \!&\! 0 \!&\! -V^{c}
\end{bmatrix} ,
\label{tGamma^a}
\\
&&\hspace{-10mm}
\underline{\tilde{\Gamma}}^{(0b)}=V^{b}\underline{1} ,
\label{tGamma^b}
\\
&&\hspace{-10mm}
\underline{\tilde{\Gamma}}^{(0c)}=
\begin{bmatrix}
-V^{b} & 0 & 0 & 0 \\
0 & V^{c} & 0 & 0 \\
0 & 0 & -V^{b} & 0 \\
0 & 0 & 0 & -V^{b}
\end{bmatrix} ,
\label{tGamma^c}
\end{eqnarray}
\end{subequations}
with
\begin{subequations}
\label{V^eo}
\begin{eqnarray}
&&\hspace{-10.5mm}
V^{a}({\bm p},{\bm p}',{\bm q})\equiv 2V_{\bm q}-\frac{V_{{\bm p}+{\bm p}'-{\bm q}}+V_{{\bm p}-{\bm p}'}}{2},
\\
&&\hspace{-10mm}
V^{b}({\bm p},{\bm p}',{\bm q})\equiv \frac{V_{{\bm p}+{\bm p}'-{\bm q}}-V_{{\bm p}-{\bm p}'}}{2},
\\
&&\hspace{-10mm}
V^{c}({\bm p},{\bm p}',{\bm q})\equiv \frac{V_{{\bm p}+{\bm p}'-{\bm q}}+V_{{\bm p}-{\bm p}'}}{2}.
\end{eqnarray}
\end{subequations}
The other quantities in eq.\ (\ref{Sigma_FLEX(p)}) may also be expressed in some convenient forms after the transformation.

Finally, we briefly consider the two-particle Green's function defined by eq.\ (\ref{calK-def}), which obeys eq.\ (\ref{calK2}).
Let us expand the quantities in eq.\ (\ref{Mat}) as eq.\ (\ref{hatGamma^(0)-exp}), where
$\underline{\chi}^{(0{\rm e})}(\vec{p},\vec{p}^{\,\prime},\vec{q})$ is obtained as
\begin{eqnarray}
&&\hspace{-6.2mm}
\langle i\alpha, i'\alpha'|\underline{\chi}^{(0{\rm e})}(\vec{p},\vec{p}^{\,\prime},\vec{q}) | j\beta, j'\beta'\rangle
\nonumber \\
&&\hspace{-10mm}
=-G_{i\alpha,j\beta}(\vec{p})G_{j'\beta',i'\alpha'}(\vec{p}_{-})\delta_{\vec{p}^{\,\prime}\vec{p}}
\nonumber \\
&&\hspace{-6.2mm}
+G_{i\alpha,3-j',\beta'}(\vec{p})G_{3-j,\beta,i'\alpha'}(\vec{p}_{-})\delta_{\vec{p}^{\,\prime},-\vec{p}+\vec{q}}.
\label{uchi^(0)(p,q)}
\end{eqnarray}
Equation (\ref{calK2}) is thereby transformed into
\begin{eqnarray}
&&\hspace{-10mm}
\underline{\cal K}(\vec{p},\vec{p}^{\,\prime},\vec{q})=\underline{\chi}^{(0{\rm e})}(\vec{p},\vec{p}^{\,\prime},\vec{q})
-\sum_{\vec{p}_{1}}\underline{\chi}^{(0)}(\vec{p},\vec{q})\underline{\Gamma}(\vec{p},\vec{p}_{1},\vec{q})
\nonumber \\
&&\hspace{10mm}
\times \underline{\chi}^{(0{\rm e})}(\vec{p}_{1},\vec{p}^{\,\prime},\vec{q}),
\label{calK-p}
\end{eqnarray}
where $\underline{\Gamma}$ is the Fourier transform of eq.\ (\ref{BS}) obeying
\begin{eqnarray}
&&\hspace{-9mm}
\underline{\Gamma}(\vec{p},\vec{p}^{\,\prime},\vec{q})=\underline{\Gamma}^{({\rm ir})}(\vec{p},\vec{p}^{\,\prime},\vec{q})-
\sum_{\vec{p}_{1}}\underline{\Gamma}^{({\rm ir})}(\vec{p},\vec{p}_{1},\vec{q})\underline{\chi}^{(0)}(\vec{p}_{1},\vec{q})
\nonumber \\
&&\hspace{10mm}
\times\underline{\Gamma}(\vec{p}_{1},\vec{p}^{\,\prime},\vec{q}).
\label{BS-p}
\end{eqnarray}
Equations (\ref{calK-p}) and (\ref{BS-p}) correspond to eqs.\ (6) and (7) of Leggett for the spin-singlet pairing
in his theory of superfluid Fermi liquids,\cite{Leggett65} respectively.
Here, we can calculate $\underline{\cal K}$ microscopically based on $\Phi$ for both the singlet and triplet pairings.

\subsection{FLEX-S approximation for the contact interaction and spin-singlet pairing}

Now, we focus on the contact interaction
\begin{equation}
V_{\bm q}=U .
\label{U-def}
\end{equation}
In this case, we can simplify eq.\ (\ref{Sigma_FLEX(p)}) for the self-energy considerably.
Indeed, substituting eq.\ (\ref{U-def}) into eq.\ (\ref{tGamma-mat}),
we observe that only the 1st, 2nd, 3rd, 4th, 10th, and 14th diagonal elements are finite in the diagonal matrix $\underline{\tilde{\Gamma}}^{(0)}$;
the relevant $6\times 6$ submatrix is given by
\begin{equation}
\underline{\check{\Gamma}}^{(0)}=U
\begin{bmatrix}
1 & 0 & 0 & 0 & 0 & 0 \\
0 & -1 & 0 & 0 & 0 & 0 \\
0 & 0 & -1 & 0 & 0 & 0 \\
0 & 0 & 0 & -1 & 0 & 0 \\
0 & 0 & 0 & 0 & 1 & 0 \\
0 & 0 & 0 & 0 & 0 & 1 
\end{bmatrix} .
\label{cGamma^(0)}
\end{equation}
Moreover, absence of momentum dependences in $\underline{\Gamma}^{(0)}$ enables us to perform the $\vec{p}_{1}$ summations 
in eqs.\ (\ref{Sigma_FLEX(p)}) and (\ref{Gamma^(1)(p)}).
Using these properties in eq.\ (\ref{Sigma_FLEX(p)}), 
we realize that we only need the 1st, 2nd, 3rd, 4th, 10th, and 14th rows and columns of the two matrices
\begin{subequations}
\label{tchi-tdchi}
\begin{equation}
\underline{\tilde{\chi}}^{(0)}(\vec{q}) \equiv \sum_{\vec{p}}
\underline{R}\underline{\chi}^{(0)}(\vec{p},\vec{q})\underline{R}^{-1},
\label{tchi-tdchi1}
\end{equation}
\begin{equation}
\underline{\delta\tilde{\chi}}^{(i\alpha,j\beta)}(\vec{p})\equiv \underline{R}\,\underline{\delta\chi}^{(i\alpha,j\beta)}(\vec{p})\underline{R}^{-1}.
\label{tchi-tdchi2}
\end{equation}
\end{subequations}
These statements on eq.\ (\ref{Sigma_FLEX(p)}) for the contact interaction hold true for an arbitrary pairing symmetry. 

We further restrict ourselves to the spin-singlet pairing with no magnetic polarization for simplicity.
The condition implies $G_{\alpha\beta}(\vec{p})=\delta_{\alpha\beta}G_{\vec{p}}$, 
$F_{\uparrow\downarrow}(\vec{p})=-F_{\downarrow\uparrow}(\vec{p})$, and 
$F_{\alpha\alpha}(\vec{p})=0$ in eq.\ (\ref{hatG(p)}).
It may also be possible for this single-component superconductivity to choose the phase of $F_{\uparrow\downarrow}(\vec{p})$ such that
its imaginary part originates solely from the $i\varepsilon_{n}$ dependence as $F_{\uparrow\downarrow}(\vec{p})=F_{\uparrow\downarrow}^{*}(\vec{p}^{\,*})$.
These relations are summarized together with eq.\ (\ref{GF-symm}) as
\begin{subequations}
\label{GF-singlet}
\begin{equation}
\begin{array}{l}
\vspace{1mm}
G_{\alpha\beta}(\vec{p})=\bar{G}_{\beta\alpha}(-\vec{p})=\delta_{\alpha\beta}G_{\vec{p}}, \\
\vspace{1mm}
F_{\uparrow\downarrow}(\vec{p})=-F_{\downarrow\uparrow}(\vec{p})=\bar{F}_{\uparrow\downarrow}(\vec{p})=-\bar{F}_{\downarrow\uparrow}(\vec{p})
\equiv F_{\vec{p}}=F_{-\vec{p}},\\
F_{\alpha\alpha}(\vec{p})=0.
\end{array} 
\end{equation}
Similarly, the elements of eq.\ (\ref{hatSigma(p)}) will obey
\begin{equation}
\hspace{-2mm}
\begin{array}{l}
\vspace{1mm}
\Sigma_{\alpha\beta}(\vec{p})=\bar{\Sigma}_{\beta\alpha}(-\vec{p})=\delta_{\alpha\beta}\Sigma_{\vec{p}}, \\
\vspace{1mm}
\Delta_{\uparrow\downarrow}(\vec{p})=-\Delta_{\downarrow\uparrow}(\vec{p})=\bar{\Delta}_{\uparrow\downarrow}(\vec{p})
=-\bar{\Delta}_{\downarrow\uparrow}(\vec{p})
\equiv \Delta_{\vec{p}}=\Delta_{-\vec{p}},\\
\Delta_{\alpha\alpha}(\vec{p})=0.
\end{array} 
\end{equation}
\end{subequations}
Substituting eq.\ (\ref{GF-singlet}) into eq.\  (\ref{DG(p)}), we observe that the $4\times 4$ matrix equation is divided into a pair of $2\times 2$ equations that are equivalent to one another. The one composed of the 1st and 4th rows  and columns are given by
\begin{equation}
\begin{bmatrix}
\vspace{1mm}
G_{\vec{p}} \!&\! F_{\vec{p}} \\
F_{\vec{p}} \!&\! -\bar{G}_{\vec{p}}
\end{bmatrix} =
\begin{bmatrix}
\vspace{1mm}
i\varepsilon_{n}-\epsilon_{\bm p}-\Sigma_{\vec{p}} \!&\! -\Delta_{\vec{p}} \\
-\Delta_{\vec{p}} \!&\! i\varepsilon_{n}+\epsilon_{\bm p}+\bar{\Sigma}_{\vec{p}}
\end{bmatrix}^{-1},
\label{DG-singlet}
\end{equation}
with $\bar{G}_{\vec{p}}\equiv G_{-\vec{p}}$ and $\bar{\Sigma}_{\vec{p}}\equiv \Sigma_{-\vec{p}}$.
The other finite submatrix from the 2nd and 3rd rows and columns are obtained from eq.\ (\ref{DG-singlet}) by changing the signs of 
$F_{\vec{p}}$ and $\Delta_{\vec{p}}$ simultaneously.

Next, we set $(i\alpha,j\beta)=(1\!\uparrow,1\!\uparrow)$, $(1\!\uparrow,2\!\downarrow)$ in eq.\ (\ref{Sigma_FLEX(p)}) to express
$\Sigma_{\vec{p}}$ and $\Delta_{\vec{p}}$ in eq.\ (\ref{DG-singlet}) in terms of $G_{\vec{p}}$ and $F_{\vec{p}}$.
Elementary calculations with eqs.\ (\ref{dchi-def})-(\ref{FF-def}), (\ref{R-def}), and (\ref{GF-singlet}) show that the relevant $6\times 6$ submatrices of eqs.\ (\ref{tchi-tdchi1}) and (\ref{tchi-tdchi2}), which are denoted by 
$\underline{\check{\chi}}^{(0)}(\vec{q})$ and $\underline{\delta\check{\chi}}^{(i\alpha,j\beta)}(\vec{p})$, respectively, are given by
\begin{subequations}
\label{check-mat}
\begin{eqnarray}
&&\hspace{-6mm}
\underline{\check{\chi}}^{(0)}
\nonumber \\
&&\hspace{-10mm}
=
\begin{bmatrix}
\vspace{1mm}
\chi^{(0)}_{-} \!&\! 0 \!&\! 0 \!&\! 0 \!&\! \sqrt{2}\chi^{(0)}_{GF} \!&\! -\sqrt{2}\chi^{(0)}_{\bar{G}F} \\
\vspace{1mm}
0 \!&\! \chi^{(0)}_{+} \!&\! 0 \!&\! 0 \!&\! 0 \!&\! 0 \\
\vspace{1mm}
0 \!&\! 0 \!&\! \chi^{(0)}_{+} \!&\! 0 \!&\! 0 \!&\! 0 \\
\vspace{1mm}
0 \!&\! 0 \!&\! 0 \!&\! \chi^{(0)}_{+} \!&\! 0 \!&\! 0 \\
\vspace{1mm}
\sqrt{2}\chi^{(0)}_{GF} \!&\! 0 \!&\! 0 \!&\! 0 \!&\! -\chi^{(0)}_{G\bar{G}} \!&\! -\chi^{(0)}_{FF} \\
-\sqrt{2}\chi^{(0)}_{\bar{G}F} \!&\! 0 \!&\! 0 \!&\! 0 \!&\! -\chi^{(0)}_{FF} \!&\! -\chi^{(0)}_{\bar{G}G} 
\end{bmatrix}\! ,
\nonumber \\
\label{cchi^(0)}
\end{eqnarray}
\begin{equation}
\underline{\delta\check{\chi}}^{(1\uparrow,1\uparrow)}=-\frac{1}{4}
\begin{bmatrix}
G \!&\! 0 \!&\! 0 \!&\! -G \!&\! \sqrt{2}F \!&\! 0 \\
0 \!&\! 2G \!&\! 0 \!&\! 0 \!&\! 0 \!&\! 0  \\
0 \!&\! 0 \!&\! 0 \!&\! 0 \!&\! 0 \!&\! 0 \\
-G \!&\! 0 \!&\! 0 \!&\! G \!&\! -\sqrt{2}F \!&\! 0 \\
\sqrt{2}F \!&\! 0 \!&\! 0 \!&\! -\sqrt{2}F \!&\! -2\bar{G} \!&\! 0 \\
0 \!&\! 0 \!&\! 0 \!&\! 0 \!&\! 0 \!&\! 0
\end{bmatrix}\! ,
\end{equation}
\begin{equation}
\underline{\delta\check{\chi}}^{(1\uparrow,2\downarrow)}=\frac{1}{4}
\begin{bmatrix}
\vspace{1mm}
F \!&\! 0 \!&\! 0 \!&\! -F \!&\! -\sqrt{2}\bar{G} \!&\! 0 \\
\vspace{1mm}
0 \!&\! -2F \!&\! 0 \!&\! 0 \!&\! 0 \!&\! 0  \\
\vspace{1mm}
0 \!&\! 0 \!&\! 0 \!&\! 0 \!&\! 0 \!&\! 0 \\
\vspace{1mm}
F \!&\! 0 \!&\! 0 \!&\! -F \!&\! -\sqrt{2}\bar{G} \!&\! 0 \\
\vspace{1mm}
0 \!&\! 0 \!&\! 0 \!&\! 0 \!&\! 0 \!&\! 0 \\
\sqrt{2}G \!&\! 0 \!&\! 0 \!&\! -\sqrt{2}G \!&\! 2F \!&\! 0 \\
\end{bmatrix}\!,
\end{equation}
\end{subequations}
where the elements of eq.\ (\ref{cchi^(0)}) are defined by
\begin{subequations}
\label{chi_GG(q)}
\begin{eqnarray}
\chi^{(0)}_{GG}(\vec{q})\!\!\!\!&\equiv&\!\!\!\! -\sum_{\vec{p}}G_{\vec{p}}\,G_{\vec{p}_{-}}=\chi^{(0)}_{GG}(-\vec{q}),
\\
\chi^{(0)}_{FF}(\vec{q})\!\!\!\!&\equiv&\!\!\!\! -\sum_{\vec{p}} F_{\vec{p}}\,F_{\vec{p}_{-}}=\chi^{(0)}_{FF}(-\vec{q}),
\\
\chi^{(0)}_{G\bar{G}}(\vec{q})\!\!\!\!&\equiv&\!\!\!\! -\sum_{\vec{p}}G_{\vec{p}}\,\bar{G}_{\vec{p}_{-}}=\chi^{(0)}_{\bar{G}G}(-\vec{q}),
\\
\chi^{(0)}_{GF}(\vec{q})\!\!\!\!&\equiv&\!\!\!\! -\sum_{\vec{p}}G_{\vec{p}}\,F_{\vec{p}_{-}}=\chi^{(0)}_{\bar{G}F}(-\vec{q}),
\\
\chi^{(0)}_{\pm}(\vec{q})\!\!\!\!&\equiv&\!\!\!\! \chi^{(0)}_{GG}(\vec{q})\pm \chi^{(0)}_{FF}(\vec{q}).
\end{eqnarray}
\end{subequations}
Let us substitute eqs.\ (\ref{cGamma^(0)}) and (\ref{check-mat}) into eq.\ (\ref{Sigma_FLEX(p)}) for $(i\alpha,j\beta)=(1\!\uparrow,1\!\uparrow)$, $(1\!\uparrow,2\!\downarrow)$. 
In both cases, the resultant Tr can be calculated separately for the two submatrices composed of the 2nd, 3rd, and 4th rows and columns and the 1st, 5th, and 6th rows and columns. 
Also using $\sum_{\vec{q}}\chi^{(0)}_{G\bar{G}}(\vec{q})\bar{G}_{\vec{p}-\vec{q}}=
\sum_{\vec{q}}\chi^{(0)}_{GG}(\vec{q})G_{\vec{p}-\vec{q}}$, etc., in the second order, we obtain
\begin{subequations}
\label{SD-FLEX-S}
\begin{eqnarray}
& &\!\! \hspace{-12mm} 
\Sigma_{\vec{p}}
=\frac{1}{2}Un-2U^{2}\sum_{\vec{q}}\chi^{(0)}_{+}(\vec{q})G_{\vec{p}-\vec{q}}
\nonumber \\
& &\!\!\! \hspace{-3mm} 
+U^{2} \sum_{\vec{q}}\biggl( \frac{3}{2}\chi^{({\rm s})}_{\vec{q}}G_{\vec{p}-\vec{q}}
+\frac{1}{2}{\rm Tr}\,\underline{\chi}^{({\rm c})}_{\vec{q}}\underline{\cal G}_{\vec{p}-\vec{q}}\biggr) ,
\label{Sigma-FLEX-S}
\end{eqnarray}
\begin{eqnarray}
& &\!\! \hspace{-12mm} 
\Delta_{\vec{p}}
=U\sum_{\vec{p}^{\,\prime}}F_{\vec{p}^{\,\prime}}
-2U^{2}\sum_{\vec{q}}\chi^{(0)}_{+}(\vec{q})F_{\vec{p}-\vec{q}}
\nonumber \\
& &\!\!\! \hspace{-3mm} 
+U^{2} \sum_{\vec{q}}\biggl( \frac{3}{2}\chi^{({\rm s})}_{\vec{q}}F_{\vec{p}-\vec{q}}
-\frac{1}{2}{\rm Tr}\,\underline{\chi}^{({\rm c})}_{\vec{q}}\underline{\cal F}_{\vec{p}-\vec{q}}
\biggr) .
\label{Delta-FLEX-S}
\end{eqnarray}
\end{subequations}
Here $n$ is the particle density
\begin{equation}
n=2\sum_{\vec{p}}G_{\vec{p}}\,{\rm e}^{i\varepsilon_{n}0_+}
\label{n}
\end{equation}
with $0_+$ an infinitesimal positive constant,\cite{LW60}
the functions $\underline{\chi}^{(0{\rm c})}_{\vec{q}}$, $\chi^{({\rm s})}_{\vec{q}}$, and $\underline{\chi}^{({\rm c})}_{\vec{q}}$ are 
defined by
\begin{subequations}
\label{chi^cs}
\begin{equation}
\underline{\chi}^{(0{\rm c})}_{\vec{q}}\equiv 
\begin{bmatrix}
\vspace{1mm}
\chi^{(0)}_{-}(\vec{q})  & \sqrt{2}\chi_{GF}^{(0)}(\vec{q}) & -\sqrt{2}\chi_{\bar{G}F}^{(0)}(\vec{q})\\
\vspace{1mm}
\sqrt{2}\chi_{GF}^{(0)}(\vec{q}) & -\chi^{(0)}_{G\bar{G}}(\vec{q}) & -\chi^{(0)}_{FF}(\vec{q}) \\
-\sqrt{2}\chi_{\bar{G}F}^{(0)}(\vec{q}) & -\chi^{(0)}_{FF}(\vec{q}) & -\chi^{(0)}_{\bar{G}G}(\vec{q}) 
\end{bmatrix},
\label{chi^(0c)}
\end{equation}
\begin{equation}
\chi^{({\rm s})}_{\vec{q}}\equiv \frac{\chi^{(0)}_{+}(\vec{q})}{1-U\chi^{(0)}_{+}(\vec{q})},
\end{equation}
\begin{equation}
\underline{\chi}^{({\rm c})}_{\vec{q}}\equiv \underline{\chi}^{(0{\rm c})}_{\vec{q}}\bigl(\underline{1}+U\underline{\chi}^{(0{\rm c})}_{\vec{q}}\bigr)^{-1},
\end{equation}
with superscripts $^{\rm c}$ and $^{\rm s}$ denoting ``charge'' and ``spin,'' respectively,
and $\underline{\cal G}_{\vec{p}}$ and $\underline{\cal F}_{\vec{p}}$ are given by
\begin{equation}
\underline{\cal G}_{\vec{p}}\equiv
\begin{bmatrix}
G_{\vec{p}} & \sqrt{2}F_{\vec{p}} & 0 \\
\sqrt{2}F_{\vec{p}}  & -2\bar{G}_{\vec{p}} & 0 \\
0 & 0 & 0
\end{bmatrix},
\end{equation}
\begin{equation}
\underline{\cal F}_{\vec{p}}=
\begin{bmatrix}
F_{\vec{p}} & -\sqrt{2}\bar{G}_{\vec{p}} & 0  \\
0 & 0 & 0 \\
\sqrt{2}G_{\vec{p}} & 2F_{\vec{p}} & 0 
\end{bmatrix},
\end{equation}
\end{subequations}
respectively.

Setting $F=0$ in eq.\ (\ref{Sigma-FLEX-S}) yields the normal-state self-energy
\begin{eqnarray}
& & \hspace{-8.5mm} 
\Sigma_{\vec{p}}^{({\rm n})}
\!=\frac{1}{2}Un+U^{2}\sum_{\vec{q}}\!\!\left[-2\chi^{(0)}_{GG}(\vec{q})G_{\vec{p}-\vec{q}}+\frac{3}{2}\frac{\chi^{(0)}_{GG}(\vec{q})G_{\vec{p}-\vec{q}}}{1-U\chi^{(0)}_{GG}(\vec{q})}\right.
\nonumber \\
& & \hspace{2.5mm} \left.
+\frac{1}{2}\frac{\chi^{(0)}_{GG}(\vec{q})G_{\vec{p}-\vec{q}}}{1+U\chi^{(0)}_{GG}(\vec{q})}
+\frac{\chi^{(0)}_{G\bar{G}}(\vec{q})\bar{G}_{\vec{p}-\vec{q}}}{1-U\chi^{(0)}_{G\bar{G}}(\vec{q})}\right] .
\label{Sigma^(n)}
\end{eqnarray}
The second term on the right-hand side completely agrees with eq.\ (2.31) by Bickers and Scalapino in the normal FLEX approximation,\cite{BS89}
as it should.
To find the equation for $T_{c}$, let us assume the second-order transition and linearize (i) the right-hand side of eq.\ (\ref{Delta-FLEX-S}) 
in terms of $F$ and (ii) $F$ itself in eq.\ (\ref{DG-singlet}) in terms of $\Delta$ as
\begin{equation}
F_{\vec{p}}\rightarrow F^{(1)}_{\vec{p}}\equiv \frac{\Delta_{\vec{p}}}{\bigl(i\varepsilon_{n}-\epsilon_{\bm p}-\Sigma^{({\rm n})}_{\vec{p}}\bigr)
\bigl(i\varepsilon_{n}+\epsilon_{\bm p}+\bar{\Sigma}^{({\rm n})}_{\vec{p}}\bigr)}.
\end{equation}
We thereby obtain the linearized self-consistency equation as
\begin{equation}
\Delta_{\vec{p}}=\sum_{\vec{p}^{\,\prime}}V^{\rm s}_{\vec{p}\vec{p}^{\,\prime}}F_{\vec{p}^{\,\prime}}^{(1)},
\end{equation}
where $V_{\vec{p}\vec{p}^{\,\prime}}^{\rm s}$ is given by
\begin{eqnarray}
&&\hspace{-10mm}
V_{\vec{p}\vec{p}^{\,\prime}}^{\rm s}
=U+U^{2}\biggl[-2\chi^{(0)}_{GG}(\vec{p}-\vec{p}^{\,\prime})
+\frac{3}{2}\frac{\chi^{(0)}_{GG}(\vec{p}-\vec{p}^{\,\prime})}{1-U\chi^{(0)}_{GG}(\vec{p}-\vec{p}^{\,\prime})}
\nonumber \\
&&\hspace{0.5mm}
-\frac{1}{2}\frac{\chi^{(0)}_{GG}(\vec{p}-\vec{p}^{\,\prime})}{1+U\chi^{(0)}_{GG}(\vec{p}-\vec{p}^{\,\prime})}
\biggr]
\nonumber \\
&&\hspace{0.5mm}
-U^{2}\sum_{\vec{q}}\frac{(G_{\vec{p}+\vec{q}}+G_{-\vec{p}+\vec{q}})G_{\vec{p}^{\,\prime}+\vec{q}}}{\bigl[1+U\chi^{(0)}_{GG}(\vec{q})\bigr]\bigl[1-U\chi^{(0)}_{G\bar{G}}(\vec{q})\bigr]}.
\label{Tc-FLEX-S-singlet}
\end{eqnarray}
This $V_{\vec{p}\vec{p}^{\,\prime}}^{\rm s}$ may be regarded as the pair potential for the spin-singlet pairing without feedback effects.\cite{AB73,Leggett75}
Note in this context that the feedback effects are included naturally in Eq.\ (\ref{Delta-FLEX-S}).

Equations (\ref{DG-singlet}), (\ref{SD-FLEX-S}), and (\ref{n}) form a closed set of self-consistent equations
for the spin-singlet pairing within the FLEX-S approximation.
The replacement
$$
\underline{\chi}^{(0{\rm c})}_{\vec{q}}\rightarrow
\begin{bmatrix}
\vspace{1mm}
\chi^{(0)}_{-}(\vec{q})  & 0 & 0 \\
\vspace{1mm}
0 & 0 & 0 \\
0 & 0 & 0
\end{bmatrix}
$$
in eq.\ (\ref{SD-FLEX-S}) yields the standard FLEX approximation for superconductivity
with only the particle-hole contribution;\cite{Tewordt74,Tewordt93,Tewordt95,PB94,MS94,GLSB96,TM98,KFY97}
the charge contribution is disregarded further in many cases.
This approximation may be appropriate for
the Hubbard model near half filling where spin fluctuations are dominant, for example.
On the other hand, neglecting the particle-hole contribution in eq.\ (\ref{SD-FLEX-S}) altogether by $\chi^{(0)}_{\pm}(\vec{q})\rightarrow 0$ and 
$$
\underline{\chi}^{(0{\rm c})}_{\vec{q}}\rightarrow  
\begin{bmatrix}
\vspace{1mm}
0  & 0 & 0 \\
\vspace{1mm}
0 & -\chi^{(0)}_{G\bar{G}}(\vec{q}) & -\chi^{(0)}_{FF}(\vec{q}) \\
0 & -\chi^{(0)}_{FF}(\vec{q}) & -\chi^{(0)}_{\bar{G}G}(\vec{q}) 
\end{bmatrix},
$$
we obtain the particle-particle ladder approximation
adopted by Haussmann {\em et al}.\ for the BCS-BEC crossover problem.\cite{Haussmann07}
Hence, eq.\ (\ref{SD-FLEX-S}) enables us to study the two cases of considerable physical interest on an equal footing,
including a crossover from the one limit to the other.
It is also clear that the FLEX+T-matrix approximation,\cite{DMT97,Yanase01} which incorporates a contribution of the two-particle Green's function into the single-particle self-energy, cannot be reproduced as a limit of the FLEX-S approximation, where both of the single-particle and two-particle Green's functions should be studied starting from the single functional of eq.\ (\ref{Phi_FLEX}). Hence, a microscopic foundation of the FLEX+T-matrix approximation remains to be examined critically by including more terms from eq.\ (\ref{Phi}) beyond the FLEX-S approximation.

\section{Summary}

We have developed a concise self-consistent perturbation expansion for superconductivity in terms of Green's function $\hat{G}$ and fully symmetrized bare vertex $\underline{\Gamma}^{(0)}$ defined by eqs.\ (\ref{hatG}) and (\ref{hatGamma^(0)}), respectively, so as to incorporate all the pair processes naturally and conveniently.
Indeed, distinct diagrams for the key functional $\Phi[\hat{G}]$ of eq.\ (\ref{Phi}) are exhausted within the fourth order by the five graphs of Fig.\ \ref{Fig5}.
Corresponding analytic expressions can be written down immediately as eq.\ (\ref{Phi_1-4}) based on the Feynman rules (a)-(f) given around eq.\ (\ref{tG_ij}).
Using an approximate $\Phi$ thereby constructed, we can calculate $\hat{G}$ self-consistently by eqs.\ (\ref{DG}) and (\ref{Sigma-Phi}), the thermodynamic potential subsequently  by eq.\ (\ref{Omega}), and also the two-particle Green's function of eq.\ (\ref{calK-def}) by eq.\  (\ref{calK}) with
eqs.\ (\ref{chi^(0)}), (\ref{Gamma^(ir)}), and (\ref{Mat}). Moreover, entropy may be written down explicitly once $\hat{G}$ and $\hat{\Sigma}$ are known.\cite{Kita99}
Thus, a single procedure of choosing an approximate $\Phi$ enables us to calculate the whole thermodynamic hierarchy, and the results may be improved by starting from a better $\Phi$. 
Various fluctuations and competing orders may be described simultaneously through the matrix structure of $\underline{\Gamma}^{(0)}$.
Since the conservation laws are obeyed automatically, the basic formalism of \S 2  in the coordinate representation can be applied to study non-equilibrium phenomena by changing the Matsubara contour into the Schwinger-Keldysh contour.\cite{Kita10}
The idea of symmetrizing the bare vertex in terms of the particle-hole indices will also be useful for the electron-phonon Hamiltonian to simplify its perturbation expansion. 

This formalism has been transformed for homogeneous systems into the momentum-energy representation in \S 3.1. We have subsequently derived a closed set of self-consistent equations for the contact potential in the spin-singlet FLEX-S approximation as eqs.\ (\ref{DG-singlet}), (\ref{SD-FLEX-S}), and (\ref{n}), which adds all the pair processes characteristic of superconductivity to the normal FLEX approximation.\cite{BS89} As shown in the paragraph after eq.\ (\ref{Tc-FLEX-S-singlet}), it contains extra terms besides those in the standard FLEX approximation for superconductivity, and also embraces the particle-particle ladder approximation by Haussmann {\em et al}.\ \cite{Haussmann07} for the BCS-BEC crossover problem.
Thus, our FLEX-S approximation incorporates the particle-hole and particle-particle diagrams on an equal footing with their mixing effects, thereby enabling us to study the BCS-BEC problem from low- to high-density regions continuously.

\begin{acknowledgments}
This work is supported by a Grant-in-Aid for Scientific Research 
from the Ministry of Education, Culture, Sports, Science and Technology of Japan.
\end{acknowledgments}

\appendix

\section{Perturbation expansion with $\Gamma^{(0)}$\label{appendix:Feynman}}

The Feynman rules for the normal-state expansion of eq.\ (\ref{Phi}) with eq.\ (\ref{S-matrix}) are summarized as follows.

\begin{enumerate}
\item[(a)] Draw all the $n$th-order closed skeleton diagrams that are topologically distinct. 
For each such diagram, associate the factor $(-1)^{n+1}/ 4^{n}n!\beta$.

\item[(b)] For each vertex, associate $\Gamma^{(0)}(\bar{\lambda}\bar{\lambda}',\bar{\nu}\bar{\nu}')$.

\item[(c)] Identify the number $C_{n\alpha}$ of possible connections of vertices for diagram $\alpha$ under consideration.

\item[(d)] Consider a specific connection of vertices for diagram $\alpha$ and associate $G(\bar{\lambda}',\bar{\eta})$ for each line arriving at $\bar{\lambda}'$  from $\bar{\eta}$.

\item[(e)] Identify the number $\ell_{n\alpha}$ of permutations to realize the connection. 
We may find it by replacing 
$\Gamma^{(0)}(11',22')\rightarrow \delta(1,1')\delta(2,2') V({\bm r}_{1}-{\bm r}_{2})$
and equating $\ell_{n\alpha}$ with the number of closed particle loops in the resultant diagram with $V$.\cite{LW60}

\item[(f)] Multiply the expression by $(-1)^{\ell_{n\alpha}}C_{n\alpha}$.

\end{enumerate}

\noindent
These rules can also be used for superconductors, 
where the number $\ell_{n\alpha}$ for an ``anomalous'' diagram
may only be found by a direct counting of permutations.

A key quantity in the rules is the numerical weight factor $C_{n\alpha}$. 
Since the two operators $\psi_{i}\psi_{i}$ in eq.\ (\ref{S-matrix}) with the same subscript $i$ ($=1,2$) are equivalent, the factor may be obtained elementarily by a combinatorial consideration. 
For example, the factors for Fig.\ \ref{Fig3}(a) and (b) are trivially identified as $C_{1a}=2$ and $C_{1b}=1$, respectively.
Together with $\ell_{1a}=2$ and  $\ell_{1b}=0$ in terms of eq.\ (\ref{GF-def}), we arrive at eq.\ (\ref{Phi_1}) for Fig.\ \ref{Fig3} following the above rules.
Similarly, the weight factors for Fig.\ \ref{Fig1} (a)-(j) are found as $C_{3a}=2^{5}2!$, $C_{3b}=2^{3}2!$, $C_{3c}=3(2^{3})^{2}$, 
$C_{3d}=3! (2^{3})^{2}$, $C_{3e}=3!  2^{5}$, $C_{3f}=3(2^{3})^{2}$, $C_{3g}=3!(2^{3})^{2}$, $C_{3h}=3! 2^{3}$, $C_{3i}=3 \cdot 2^{5}$, and 
$C_{3j}=2^{6}$, each of which corresponds to the number of possible connections of three vertices for realizing the relevant topological structure. We thereby obtain the relative weights of Fig.\ \ref{Fig1}.
 
\section{Momentum-energy representation for lattice models\label{appendix:Hubbard}}

Here, we discuss the momentum-energy representation of homogeneous lattice models.
First of all, momentum integrations in eqs.\ (\ref{GS-exp}) and (\ref{V-exp}) should be restricted to the first Brillouin zone.
This fact may be incorporated manifestly in these formulas by introducing the function
\begin{equation}
\Theta_{{\bm p}}\equiv \int_{1^{\rm st}{\rm BZ}} \frac{{\rm d}^{3}k}{(2\pi)^{3}}\delta^{3}({\bm k}-{\bm p})
\label{Theta}
\end{equation}
on their right-hand sides.

With these trivial modifications, we consider the basic two-particle quantities in eq.\ (\ref{uGamma-uchi}).
Substituting eq.\ (\ref{Gamma^(0)}) into eq.\ (\ref{hatGamma^(0)-2}), we can write the matrix element of eq.\ (\ref{uGamma^(0)}) 
explicitly in terms of $V$ as
\begin{eqnarray}
&&\hspace{-4mm}
\langle 11'_{ii'}| \underline{\Gamma}^{(0)}|22'_{jj'}\rangle
\nonumber \\
&&\hspace{-8mm}
=
\frac{1}{2}\bigl[(-1)^{i+j}\delta_{ii'}\delta_{jj'}V({\bf r}_{1}'-{\bf r}_{2}')\delta(\tau_{1}'-\tau_{2}')
\delta(1,1')\delta(2',2)
\nonumber \\
&&\hspace{-4mm}
-(-1)^{i+i'}\delta_{ij}\delta_{i'j'}V({\bf r}_{2}-{\bf r}_{2}')\delta(\tau_{2}-\tau_{2}')
\delta(1,2)\delta(2',1')
\nonumber \\
&&\hspace{-4mm}
+(-1)^{i+i'}\delta_{j',3-i}\delta_{j,3-i'}
V({\bf r}_{2}'-{\bf r}_{1}')\delta(\tau_{2}'-\tau_{1}')\delta(1,2')
\nonumber \\
&&\hspace{-4mm}
\times \delta(1',2)\bigr].
\label{uGamma^(0)-exp-a}
\end{eqnarray}
Let us substitute eqs.\ (\ref{delta-fn}) and (\ref{V-exp}) with eq.\ (\ref{Theta}) into eq.\ (\ref{uGamma^(0)-exp-a})
and transform the resultant expression into the form of eq.\ (\ref{hatGamma^(0)-exp}) by changing integration variables.
We thereby obtain the Fourier coefficient in eq.\ (\ref{hatGamma^(0)-exp}) as
\begin{eqnarray}
&&\hspace{-6mm}
\langle i\alpha, i'\alpha'|\underline{\Gamma}^{(0)}({\bm p},{\bm p}^{\prime},{\bm q}) | j\beta, j'\beta'\rangle
\nonumber \\
&&\hspace{-10mm}
=\frac{1}{2}\bigl[
(-1)^{i+j}\delta_{ii'}\delta_{jj'}\delta_{\alpha'\alpha}\delta_{\beta'\beta}
\Theta_{{\bm p}}\Theta_{{\bm p}'}\Theta_{{\bm q}}V_{{\bm q}}
\nonumber \\
&&\hspace{-6mm}
-(-1)^{i+i'}\delta_{ij}\delta_{i'j'}\delta_{\alpha\beta}\delta_{\alpha'\beta'}
\Theta_{{\bm p}}\Theta_{{\bm p}-{\bm q}}\Theta_{{\bm p}-{\bm p}'}V_{{\bm p}-{\bm p}'}
\nonumber \\
&&\hspace{-6mm}
+(-1)^{i+i'}\delta_{j',3-i}\delta_{j,3-i'}\delta_{\alpha\beta'}\delta_{\alpha'\beta}
\Theta_{{\bm p}}\Theta_{{\bm p}'}\Theta_{{\bm p}+{\bm p}'-{\bm q}} 
\nonumber \\
&&\hspace{-6mm}
\times
V_{{\bm p}+{\bm p}'-{\bm q}} \bigr] .
\label{Gamma-p-a}
\end{eqnarray}
Note that ${\bm p}^{\prime}$ and ${\bm q}$ above may lie outside the first Brillouin zone to give a finite value for the matrix element.
Similarly, the Fourier coefficient of eq.\ (\ref{hatChi^(0)-exp}) for lattice models is obtained as
\begin{eqnarray}
&&\hspace{-8mm}
\langle i\alpha, i'\alpha'|\underline{\chi}^{(0)}(\vec{p},\vec{q}) | j\beta, j'\beta'\rangle
=-G_{i\alpha,j\beta}(\vec{p}\,)G_{j'\beta',i'\alpha'}(\vec{p}_{-}) 
\nonumber \\
&&\hspace{37mm}\times
\Theta_{{\bm p}}\Theta_{{\bm p}-{\bm q}}.
\label{chi^(0)(p,q)-a}
\end{eqnarray}
By incorporating $\Theta$ functions accordingly in other formulas, the whole results of \S 3 also become applicable to lattice models.

\end{document}